\documentclass{article}

\usepackage{graphicx}
\usepackage{wasysym}

\def\eref#1{(\ref{eq:#1})}
\def\rfig#1{fig.~\ref{fig:#1}}

\def\beq{\begin{equation}}
\def\eeq{\end{equation}}
\def\comment#1{\noindent{\bf }}
\newcommand{\dpar}[2]{{\partial #1\over\partial #2}}

\def\bfnab{{\bf\nabla}}
\def\bfOm{{\bf\Omega}}
\def\bfr{{\bf r}}
\def\ie{{\it i.e.}}
\def\etal{{\it et al.}}
\def\rbar{{\bar{r}}}

\def\ind#1{{$^{#1}$}}
\def\wig#1{\mathrel{\hbox{\hbox to 0pt{%
          \lower.6ex\hbox{$\sim$}\hss}\raise.4ex\hbox{$#1$}}}}

\def\req{R_{\rm eq}}
\def\rpol{R_{\rm pol}}

\def\mea{{\rm\,M_\oplus}}
\def\msol{{\rm\,M_\odot}}
\def\mjup{{\rm\,M_J}}

\def\eg{{\it e.g.}}

\def\linc{L_{\rm *p}}
\def\teff{T_{\rm eff}}
\def\teq{T_{\rm eq}}

\def\g{\rm\,g}

\def\K{\rm\,K}
\def\gcc{\rm\,g\,cm^{-3}}

\def\erg{\rm\, erg}
\def\p#1{\times 10^{#1}}
\def\ti#1{$^{\rm #1}$}

\def\ind{\par\noindent\hangindent=2.5em\hangafter=1}
\def\bib{\par\noindent\hangindent=2.5em\hangafter=1}
\newcommand{\artt}[6]{\ind{{#1}. #6. #2. {\it #3\/} {#4}:#5}}

\newcommand{\boa}[4]{\ind{{#1}. #4. In {\it #2}, #3}}
\newcommand{\boo}[4]{\ind{{#1}. #4. {\it ``#2''}, #3}}

\def\ApJ{ApJ}
\def\apj{ApJ}
\def\apjs{ApJS}
\def\apjl{ApJL}

\def\nat{Nature}

\def\icarus{Icarus}

\def\JGR{J. Geophys. Res.}

\def\pss{Plan. Space. Sci.}
\def\planss{Plan. Space. Sci.}

\begin{document}

\title{\bf The Interiors of Giant Planets \\
Models and Outstanding Questions}
\author{{\sc Tristan Guillot},\\
\normalsize Observatoire de la C\^ote d'Azur, \\
\normalsize Laboratoire Cassiop\'ee, CNRS UMR 6202, \\
\normalsize 06304 Nice Cedex 04, France\\
\normalsize guillot@obs-nice.fr
}
\date{To appear in Ann. Rev. Earth \& Plan. Sciences (2005)}
%Key words: giant planets, extrasolar planets, planet formation
\maketitle

\begin{abstract} 
We know that giant planets played a crucial role in the making of our
Solar System. The discovery of giant planets orbiting other stars is a
formidable opportunity to learn more about these objects, what is
their composition, how various processes influence their structure and
evolution, and most importantly how they form.  Jupiter, Saturn,
Uranus and Neptune can be studied in detail, mostly from close
spacecraft flybys. We can infer that they are all enriched in heavy
elements compared to the Sun, with the relative global enrichments
increasing with distance to the Sun. We can also infer that they
possess dense cores of varied masses. The intercomparison of presently
caracterised extrasolar giant planets show that they are also mainly
made of hydrogen and helium, but that they either have significantly
different amounts of heavy elements, or have had different orbital
evolutions, or both. Hence, many questions remain and are to be
answered for significant progresses on the origins of planets.
\end{abstract}

%----------------------------------------------------------
%\tableofcontents
%----------------------------------------------------------
\newpage
\begin{flushright}
\it
---Pourquoi l'azur muet et l'espace insondable?\\
Pourquoi les astres d'or fourmillant comme un sable?\\
\rm
Arthur Rimbaud---Soleil et chair
\end{flushright}

\section{Introduction}

Looking at a starry sky, it is quite vertiginous to think that we are
at one of these special epochs in history: Just before, we only knew
of the planets in our Solar System. Now, more than 150
giant planets are known to orbit solar-like stars. Our
giant planets, Jupiter, Saturn, Uranus and Neptune are no longer the
only ones that we can characterize. We now know of six extrasolar
giant planets transiting in front of their stars, for which we can
measure with a fair accuracy their mass and radius. We lie on
the verge of a true revolution: With ground-based and future
space-based transit search programs, we should soon be able to detect
and characterize many tens, probably hundreds of planets orbiting
their stars, with the hope of inferring their composition and hence
the mechanisms responsible for the formation of planets. 

It is a daunting task too, because we should expect that, like for the
planets in our Solar System, a rich variety of giant planets is found,
with different compositions, different histories and a number of
new or unexpected physical mecanisms at work. We will have to
classify observations, test theories, and be aware that although
simplicity is appealing, it is not always what Nature has in store for
us. 

This review aims at providing a synthetic approach to the problems
posed by ``old'' and ``new'' giant planets, in the Solar System and
outside. It updates a previous review by Stevenson (1982), and expands
on the review by Hubbard et al. (2002) by focusing on the mass-radius
relations and compositions of giant planets.  In Section~2, we see how
to construct interior and evolution models of giant planets. Section~3
is devoted to our giant planets, what we can infer from observations,
and the questions that remain.  I then turn to the new
field of extrasolar giant planets, focusing on the close-in,
transiting ``Pegasi planets'' (also called ``hot Jupiters''). The last
section is an attempt to summarize some of the known facts concerning
giant planets and provide a few expected milestones for future
studies.

\section{The calculation of interior models}

\subsection{A simple model}

To tell our story, I will use a simple model, based on the following
assumptions:
\begin{enumerate}
\item Giant planets are made of a fluid envelope and possibly a
  dense central core of about $\sim 15\rm\,M_\oplus$ (Earth masses);
\item The envelope is mostly made of hydrogen and helium and trace
  species (heavy elements); The core is made of an unknown combination
  of refractory material (``rocks'') and more volatile species
  (``ices'' including molecular species such as H2O water, CH4 methane
  and NH3 ammonia in the fluid state);
\item Contrary to solid planets, viscosity is negligible
  throughout;
\item In most cases, rotation and magnetic fields can be neglected;
\item Giant planets were formed from an extended, high-entropy, 
  high-luminosity state.
\end{enumerate}
These assumptions can only be justified {\it a posteriori}: they are
the result of our knowledge of observed giant planets and of
inferences about the mechanisms that led to their formation.  We will
see how this simple model predicts the global properties of giant
planets between 1/20 to 20$\mjup$ (about 15 and 6000\,M$_\oplus$), and
how they compare with observations.

\subsection{Basic equations}

As a consequence of our assumptions, the structure and evolution of a
giant planet is governed by the following hydrostatic, thermodynamic,
mass conservation and energy conservation equations:
\begin{eqnarray}
\dpar{P}{r}&=&-\rho g \label{eq:dpdr}\\
{\partial T\over\partial r}&=&{\partial P\over \partial r}{T\over
P}\nabla_T. \label{eq:dtdr}\\
{\partial m\over\partial r}&=&4\pi r^2\rho. \label{eq:dmdr}\\
{\partial L\over\partial r}&=&4\pi r^2\rho \left(\dot{\epsilon}-
T{\partial S\over \partial t}\right),\label{eq:dldr}
\end{eqnarray}
where $P$ is the pressure, $\rho$ the density, and $g=Gm/r^2$ the
gravity ($m$ is the mass, $r$ the radius and $G$ the gravitational
constant). The temperature gradient $\nabla_T\equiv(d\ln T/d\ln P)$
depends on the process by which the internal heat is transported.
$L$ is the intrinsic luminosity, $t$ the time, $S$ the
specific entropy (per unit mass), and $\dot{\epsilon}$ accounts for
the sources of energy due e.g. to radioactivity or more importantly
nuclear reactions. Generally it is a good approximation to assume
$\dot{\epsilon}\sim 0$ for objects less massive than $\sim 13\mjup$,
i.e. too cold to even burn deuterium (but we will see that in certain
conditions this term may be useful, even for low mass planets). 

The boundary condition at the center is trivial: $r=0$; ($m=0$,
$L=0$). The external boundary condition is more difficult to obtain
because it depends on how energy is transported in the atmosphere. One
possibility is to use the Eddington approximation, and to write (\eg\
Chandrasekhar 1939): $r=R$; ($T_0=\teff$, $P_0=2/3\,g/\kappa$), where
$\teff$ is the effective temperature, and $\kappa$ is the opacity in
$\rm cm^2\,g^{-1}$. Note for example that in the case of Jupiter
$\teff=124$\,K, $g=2600\rm\,cm\,s^{-2}$ and $\kappa\approx 5\times
10^{-2} (P/1\rm\,bar)\,cm^2\,g^{-1}$. This implies $P_0\approx
0.2$\,bar, which is actually close to Jupiter's tropopause, where
$T\approx 110$\,K.

More generally, one has to use an atmospheric model relating the
temperature and pressure at a given level to the radius $R$, intrinsic
luminosity $L$ and incoming stellar luminosity $\linc$: $r=R$;
($T_0=T_0(R,L,\linc)$, $P_0=P_0(R,L,\linc)$). $P_0$ is chosen to
satisfy the condition that the corresponding optical depth at that
level should be much larger than unity. If the stellar flux is
absorbed mostly in a convective zone, then the problem can be
simplified by using $T_0(R,L,\linc)\approx T_0(R,L+\linc,0)$
(e.g. Hubbard 1977). An example of such a model is described by Saumon
et al. (1996) and Hubbard et al. (2002) and is used hereafter to model
the planets in the low irradiation limit.

\subsection{High pressure physics \& equations of state}

\begin{figure}[htbp]
\hspace*{-2cm}
\resizebox{16cm}{!}{\includegraphics[angle=90]{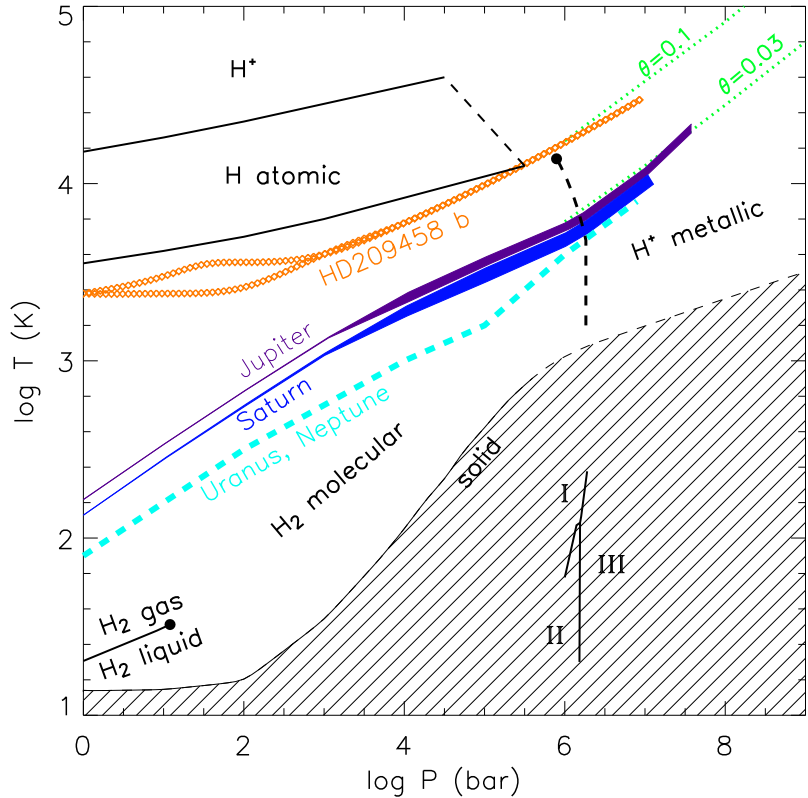}}
%\centerline{\hbox{\psfig{file=diag_phase.ps,width=14cm}}}
\caption{Phase diagram for hydrogen with the main phase transitions
occurring in the fluid or gas phase. The temperature-pressure profiles
for Jupiter, Saturn, Uranus, Neptune, and HD209458 b are shown. The
dashed nearly vertical line near 1\,Mbar is indicative of the
molecular to metallic transition (here it represents the so-called
plasma phase transition as calculated by Saumon et al. 2000). The
region in which hydrogen is in solid phase (Datchi et al. 2000;
Gregoryanz et al. 2003) is represented as a dashed area. The three
phases (I,II,III) of solid hydrogen are shown (see Mao \& Hemley
1994). Values of the degeneracy parameter $\theta$ are indicated as
dotted lines to the upper right corner of the figure.}
\label{fig:phase_diag}
\end{figure}

In terms of pressures and temperatures, the interiors of giant planets
lie in a region for which accurate equations of state (EOS) are
extremely difficult to calculate. This is because both molecules,
atoms and ions can coexist, in a fluid that is partially degenerate
(free electrons have energies that are determined both by quantic and
thermal effects) and partially coupled (coulombian interactions
between ions are not dominant but must be taken into account). The
presence of many elements and their possible interactions further
complicate matters. For lack of space, this section will mostly focus
on hydrogen whose EOS has seen the most important developments in
recent years. A phase diagram of hydrogen (fig.~\ref{fig:phase_diag})
illustrates some of the important phenomena that occur in giant
planets.

The photospheres of giant planets are generally relatively cold (50 to
3000\,K) and at low pressure (0.1 to 10\,bar, or $10^4$ to
$10^6$\,Pa), so that hydrogen is in molecular form and the perfect gas
conditions apply.  As one goes deeper into the interior hydrogen and
helium progressively become fluid. (The perfect gas relation tends to
underestimate the pressure by 10\% or more when the density becomes
larger than about $0.02\gcc$ ($P\wig{>} 1\,$kbar in the case of
Jupiter)).

Characteristic interior pressures are considerably larger however: as
implied by Eqs.~\ref{eq:dpdr} and \ref{eq:dmdr}, $P_{\rm c}\approx
GM^2/R^4$, of the order of 10-100\,Mbar for Jupiter and Saturn. At
these pressures and the corresponding densities, the Fermi temperature
$T_{\rm F}$ is larger than $10^5$\,K. This implies that electrons are
degenerate. Figure~\ref{fig:phase_diag} shows that inside Jupiter,
Saturn, HD209458b, but also for giant planets in general for most of
their history, the degeneracy parameter $\theta=T/T_{\rm F}$ is
between 0.1 and 0.03. Therefore, the energy of electrons in the
interior is expected to be only slightly larger than their
non-relativistic, fully degenerate limit: $u_{\rm e}\ge 3/5\,kT_{\rm
F} =15.6\left(\rho/\mu_{\rm e}\right)^{2/3}\ \rm eV$, where $k$ is
Boltzmann's constant, $\mu_{\rm e}$ is the number of electrons per
nucleon and $\rho$ is the density in $\rm g\,cm^{-3}$. For pure
hydrogen, when the density reaches $\sim 0.8\gcc$, the average energy
of electrons becomes larger than hydrogen's ionization potential, even
at zero temperature: hydrogen pressure-ionizes and becomes
metallic. This molecular to metallic transition occurs near Mbar
pressures, but exactly how this happens remains unclear because of the
complex interplay of thermal, coulombian, and degeneracy effects (in
particular, whether hydrogen metallizes into an atomic state H$^+$ ---
as suggested in Fig. 1 --- or first metallizes in the molecular state
H$_2$ remains to be clarified).

Recent laboratory measurements on fluid deuterium have been able to
reach pressures above $\wig> 1\,$Mbar, and provide new data in a
region where the EOS remains most uncertain. Gas-guns experiments have
been able to measure the reshock temperature (Holmes et al. 1995),
near $T\sim 5000\K$, $P\sim 0.8\,$Mbar, and a rise in the conductivity
of molecular hydrogen up to $T\sim 3000\K$, $P\sim 1.4\,$Mbar, a sign
that metallicity may have been reached (Weir et al. 1996). The following few
years have seen the development of laser-induced shock compression (Da
Silva et al. 1997, Collins et al. 1998), pulsed-power shock
compression (Knudson et al. 2002, 2004), and convergent shock wave
experiments (Belov et al. 2002; Boriskov et al. 2003) in a
high-pressure ($P=0.3-4\,$Mbar) high-temperature ($T\sim
6000-10^5\K$) regime. Unfortunately, experimental results along the
principal Hugoniot of deuterium do not agree in this pressure range.
Laser compression data give a maximum
compression of $\sim 6\,$ while both the pulsed-power compression
experiments and the convergent shock wave experiments find a value of
$\sim 4\,$. Models that are partly calibrated with experimental data
(Saumon, Chabrier \& Van Horn 1995; Ross 1998; Saumon et al. 2000,
Ross \& Yang 2001) obtain a generally good agreement with the
laser-compression data. However, the fact that independant models
based on first principles (Militzer \& Ceperley 2001; Desjarlais 2003;
Bonev et al. 2004) yield low compressions strongly favors this
solution. 

The question of the existence of a first-order molecular to metallic
transition of hydrogen (i.e. both molecular dissociation and
ionisation occur simultaneously and discontinuously at the so-called
plasma phase transition, or PPT) remains however. The critical line
shown in \rfig{phase_diag} corresponds to calculations by Saumon et
al. (2000), but may be caused by artefacts in the free energy
calculation. Recent Density Functional Theory (DFT) simulations by
Bonev et al. (2004) indicate the possibility of a first order
liquid-liquid transition but other path-integral calculations
(Militzer \& Ceperley 2001) do not. It is crucial to assess the
existence of such a PPT because it would affect both convection and
chemical composition in the giant planets.

A clear result from \rfig{phase_diag} at least is that, as first shown
by Hubbard (1968), the interiors of the hydrogen-helium giant planets
are {\it fluid}, whatever their age (an {\it isolated} Jupiter should
begin partial solidification only after at least $\sim 10^3\,$Ga of
evolution). For Uranus and Neptune, the situation is actually more
complex because at large pressures they are not expected to contain
hydrogen, but numerical simulations show that ices in their interior
should be fluid as well (Cavazzoni et al. 1999).

Models of the interiors of giant planets require thermodynamically
consistent EOSs calculated over the entire domain of pressure and
temperature spanned by the planets during their evolution. Elements
other than hydrogen, most importantly helium, should be consistently
included. Such a calculation is a daunting task, and the only recent
attempt at such an astrophysical EOS for substellar objects is that by
Saumon et al. (1995). Another set of EOSs reproducing either the high-
or low-compression results was calculated by Saumon \& Guillot (2004)
specifically for the calculation of present-day models of Jupiter and
Saturn.

These EOSs have so far included other elements (including helium),
only in a very approximative way, i.e. with EOSs for helium and heavy
elements that are based on interpolations between somewhat ideal
regimes, using an additive volume law, and neglecting the possibility
of existence of phase separations (see Hubbard et al. 2002 and Guillot
et al. 2004 for further discussions).

\subsection{Heat transport}

Giant planets possess hot interiors, implying that a
relatively large amount of energy has to be transported from the deep
regions of the planets to their surface. This can either be done by
radiation, conduction, or, if these processes are not sufficient, by
convection. Convection is generally ensured by the rapid rise of the
opacity with increasing pressure and temperature.  At pressures of a
bar or more and relatively low temperatures (less than 1000\,K), the
three dominant sources of opacities are water, methane and
collision-induced absorption by hydrogen molecules.

However, in the intermediate temperature range between $\sim 1200$ and
$1500\K$, the Rosseland opacity due to the hydrogen and helium
absorption behaves differently: the absorption at any given wavelength
increases with density, but because the temperature also rises, the
photons are emitted at shorter wavelengths, where the monochromatic
absorption is smaller. As a consequence, the opacity can
decrease. This was shown by Guillot et al. (1994) to potentially lead
to the presence of a deep radiative zone in the interiors of Jupiter,
Saturn and Uranus.

This problem must however be reanalyzed in the light of recent
observations and analyses of brown dwarfs. Their spectra show
unexpectedly wide sodium and potassium absorption lines (see Burrows,
Marley \& Sharp 2000), in spectral regions where hydrogen, helium,
water, methane and ammonia are relatively transparent. It thus appears
that the added contribution of these elements (if they are indeed
present) would wipe out any radiative region at these
levels (Guillot et al. 2004). 

At temperatures above $1500\sim 2000\K$ two important sources of
opacity appear: (i) the rising number of electrons greatly enhances
the absorption of H$_2^-$ and H$^-$; (ii) TiO, a very strong absorber
at visible wavelengths is freed by the vaporization of
CaTiO$_3$. Again, the opacity rises rapidly which ensures a convective
transport of the heat. Still deeper, conduction by free electrons
becomes more efficient, but the densities are found not to be high
enough for this process to be significant, except perhaps near the
central core (see Hubbard 1968; Stevenson \& Salpeter 1977).

However, because irradiated giant planets do develop a radiative zone,
Rosseland opacity tables covering the proper range of temperatures and
pressures are needed. A pure hydrogen-helium mixture table has been
calculated by Lenzuni et al. (1991). Opacities for solar composition
including grains are available from Alexander \&
Ferguson (1994), but they do not include alkali metals and up-to-date
data on water, methane and TiO absorption. Guillot (1999a) provides a
grain-free, alkali-free table which is limited to low-temperature
regimes. The calculations hereafter use opacities provided by
F. Allard on the basis of calculations for brown dwarfs of solar
composition, including grains and alkali metals (Allard et al. 2001).

\subsection{The contraction and cooling histories of giant planets}

The interiors of giant planets is expected to evolve with time from a
high entropy, high $\theta$ value, hot initial state to a low entropy,
low $\theta$, cold degenerate state. The essential physics behind can
be derived from the well-known virial theorem and the energy
conservation which link the planet's internal energy $E_{\rm i}$,
gravitational energy $E_{\rm g}$ and luminosity through:
\begin{eqnarray}
\xi E_{\rm i} + E_{\rm g} &=&0,\\
L &=& -{\xi-1\over \xi}{dE_{\rm g}\over dt},
\end{eqnarray}
where $\xi=\int_0^M 3(P/\rho)dm / \int_0^M u dm\approx <\!3P/\rho u\!>$
and $u$ is the specific internal energy. For a diatomic perfect gas,
$\xi=3.2$; for fully-degenerate non-relativistic electrons, $\xi=2$. 

Thus, for a giant planet or brown dwarf beginning its life mostly as a
perfect H$_2$ gas, two third of the energy gained by contraction is
radiated away, one third being used to increase $E_{\rm i}$. The
internal energy being proportional to the temperature, the effect is
to heat up the planet. This represents the slightly counter-intuitive
but well known effect that a star or giant planet initially heats up
while radiating a significant luminosity.

Let us now move further in the evolution, when the contraction has
proceeded to a point where the electrons have become degenerate.  For
simplicity, I will ignore Coulombian interactions and exchange terms,
and assume that the internal energy can be written as $E_{\rm
i}=E_{\rm el}+E_{\rm ion}$, and that furthermore $E_{\rm el}\gg E_{\rm
ion}$ ($\theta$ is small). Because $\xi\approx 2$, we know that half of
the gravitational potential energy is radiated away and half of it
goes into internal energy.  The problem is to decide how this energy
is split into an electronic and an ionic part.  The gravitational
energy changes with some average value of the interior density as
$E_{\rm g}\propto 1/R \propto \rho^{1/3}$. The energy of the
degenerate electrons is essentially the Fermi energy: $E_{\rm
el}\propto \rho^{2/3}$. Therefore, $\dot{E}_{\rm el}\approx 2(E_{\rm
e}/ E_{\rm g})\dot{E}_{\rm g}$. Using the virial theorem, this yields:
\begin{eqnarray}
\dot{E}_{\rm e}&\approx& -\dot{E}_{\rm g}\approx 2L \\
L &\approx& -\dot{E}_{\rm ion} \propto -\dot{T}.  
\end{eqnarray} 
The gravitational energy lost is entirely absorbed by the degenerate
electrons, and the observed luminosity is due to the thermal cooling
of the ions. 

Several simplifications limit the applicability of this result (that
would be valid in the white dwarf regime). In particular, the
coulombian and exchange terms in the EOS introduce negative
contributions that cannot be neglected. However, the approach is
useful to grasp how the evolution proceeds: in its
very early stages, the planet is very compressible. It follows a
standard Kelvin-Helmoltz contraction. When degeneracy sets in, the
compressibility becomes much smaller ($\alpha T\sim 0.1$, where
$\alpha$ is the coefficient of thermal expansion), and the planet
gets its luminosity mostly from the thermal cooling of the ions. The
luminosity can be written in terms of a modified Kelvin-Helmoltz
formula: 
\begin{equation}
L\approx \eta {GM^2\over R\tau},
\label{eq:lapprox}
\end{equation}
where $\tau$ is the age, and $\eta$ is a factor that hides most of the
complex physics. In the approximation that Coulombian and exchange
terms can be neglected, $\eta\approx\theta/(\theta +1)$. The poor
compressibility of giant planets in their mature evolution stages
imply that $\eta\ll 1$: the luminosity is not obtained from the entire
gravitational potential, but from the much more limited reservoir
constituted by the thermal internal energy. Equation~\ref{eq:lapprox}
shows that to first order, $\log L\propto -\log\tau$: very little time
is spent at high luminosity values. In other words, the problem is (in
most cases) weakly sensitive to initial conditions.

\begin{figure}[htbp]
\resizebox{12cm}{!}{\includegraphics[angle=0]{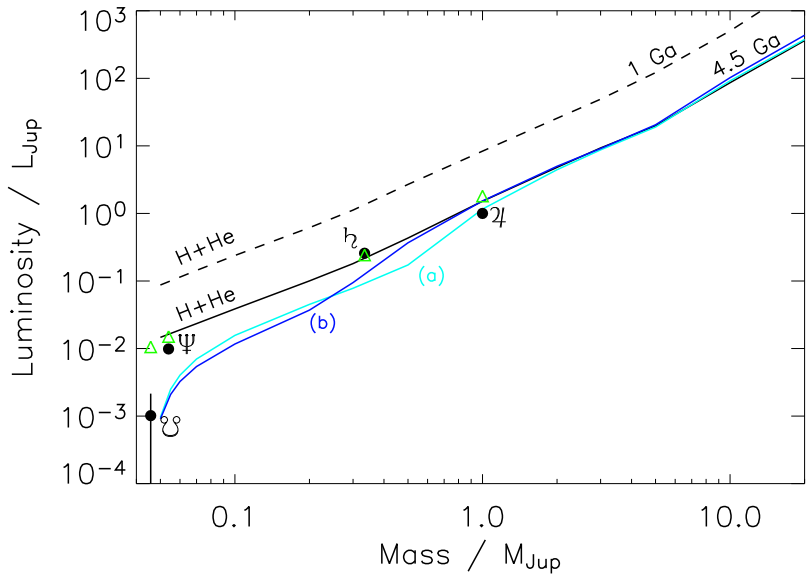}}
%\centerline{\hbox{\psfig{file=massradius.ps,width=12cm}}}
\caption{Luminosity versus mass for giant planets after
  4.5\,Ga of evolution compared to measured values for our four giant
  planets (including the significant uncertainty on Uranus'
  luminosity). The lines correspond to: H+He: a pure 
  hydrogen-helium composition with a helium mass mixing ratio
  $Y=0.25$; (a): a model with $Y=0.30$
  and a $15\mea$ core; (b): the same model but with $Y=0.36$.}
\label{fig:m-l}
\end{figure}

Figure~\ref{fig:m-l} shows calculated luminosities in the framework of
our simple model. Compared to \eref{lapprox}, calculated luminosities
are consistent with $\eta\approx 0.01$ to $0.03$. The lower
luminosities obtained in the presence of a core and of more heavy
elements are due to an earlier contraction, and quicker loss of the
internal heat. As model (b) would be appropropriate to explain
Saturn's radius (see next section), it can be seen that the planet
emits more heat than predicted by homogeneous contraction models. The
cases of Uranus and Neptune is more complex and cannot be directly
compared with the models in \rfig{m-l} which neglect the thermal heat
content of the central core.

\subsection{Mass-radius relation}

\begin{figure}[htbp]
\resizebox{12cm}{!}{\includegraphics[angle=0]{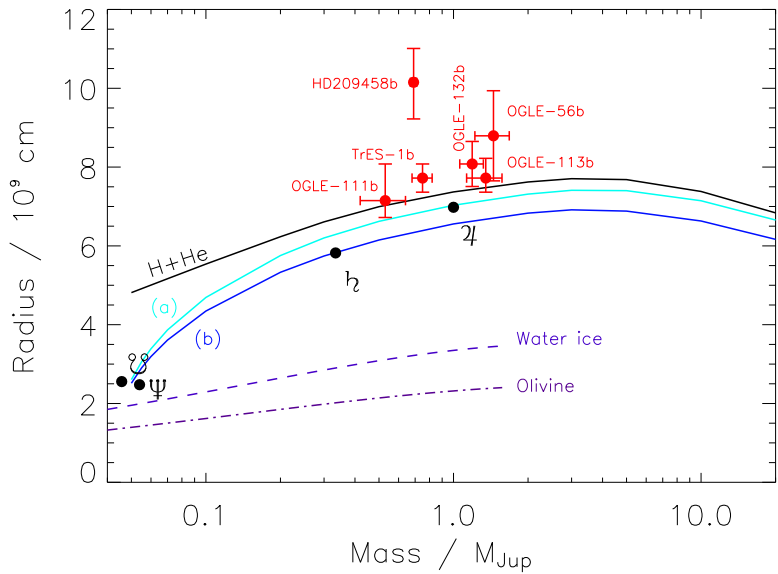}}
%\centerline{\hbox{\psfig{file=massradius.ps,width=12cm}}}
\caption{Radius versus mass for giant planets after
4.5\,Ga of evolution compared to measured values for our four giant
planets and four known extrasolar planets. As  in \rfig{m-l}, the 
lines correspond to: H+He: a pure, $Y=0.25$, hydrogen-helium composition
(Y=0.25); (a): a model with $Y=0.30$ and a $15\mea$ core; (b):
the same model but with $Y=0.36$. An approximate mass-radius relation for
zero-temperature water and olivine planets is shown as dashed and
dash-dotted lines, respectively (Courtesy of W.B. Hubbard).}
\label{fig:m-r}
\end{figure}

The relation between mass and radius has very fundamental astrophysical
applications. Most importantly it allows one to infer the gross
composition of an object from a measurement of its mass and
radius. This is especially relevant in the context of the discovery 
of extrasolar planets with both radial velocimetry and the transit
method, as the two techniques yield relatively accurate determination
of $M$ and $R$. 

Figure~\ref{fig:m-r} shows the mass-radius relation for isolated or
nearly-isolated gaseous planets, based on our simple simple model and
various assumption on their composition. The curves have a local
maximum near $4\mjup$: at small masses, the
compression is rather small so that the radius increases with mass. At
large masses, degeneracy sets in and the radius decreases with mass.

This can be understood on the basis of polytropic models based on the
assumption that $P=K\rho^{1+1/n}$, where $K$ and $n$ are
constants. Because of degeneracy, a planet of large mass will tend to
have $n\rightarrow 1.5$, while a planet a smaller mass will be less
compressible ($n\rightarrow 0$). Indeed, it can be shown that in their
inner 70 to 80\% in radius isolated planets of 10, 1 and $0.1\mjup$
have $n=1.3$, 1.0 and 0.6, respectively. From polytropic equations
(e.g. Chandrasekhar 1939):
\begin{equation}
R\propto K^{n\over 3-n} M^{1-n\over 3-n}.
\label{eq:m-r-k}
\end{equation}
Assuming that $K$ is independant of mass, one gets $R\propto
M^{0.16}$, $M^{0}$, and $M^{-0.18}$ for $M=10$,
1 and $0.1\mjup$, respectively, in relatively good agreement with
\rfig{m-r} (the small discrepancies are due to the fact that the
intrinsic luminosity and hence $K$ depend on the mass considered).

Figure~\ref{fig:m-r} shows already that the planets in our Solar
System are not made of pure hydrogen and helium: their radii lie below
that predicted for $Y=0.25$ objects. Indeed, Jupiter, Saturn, and the
two ice-giants Uranus and Neptune contain a growing proportion of
heavy elements. The theoretical curves for olivine and ice planets
predict even smaller radii however: even Uranus and Neptune contain 10
to 20\% of their mass as hydrogen and helium.

The extrasolar planets detected so far (see table~\ref{tab:transits}
hereafter) all lie above the pure hydrogen-helium curve. This is due
to the fact that these planets have their evolutions dominated by the
intense stellar irradiation they receive. Thermal effects are no
longer negligible: Using the Eddington approximation, assuming
$\kappa\propto P$ and a perfect gas relation in the atmosphere, one
can show that $K\propto (M/R^2)^{-1/2n}$ and that therefore $R\propto
M^{1/2-n\over 2-n}$. With $n=1$, one finds $R\propto
M^{-1/2}$. Strongly irradiated hydrogen-helium planets of small masses
are hence expected to have the largest radii which qualitatively
explain the positions of the extrasolar planets in \rfig{m-r}. 
Note that this estimate implicitly assumes that $n$ is constant
throughout the planet. The real situation is more complex because of
the growth of a deep radiative region in most irradiated planets, and
because of structural changes between the degenerate interior and the
perfect gas atmosphere.

\subsection{Rotation and the figures of planets}
\label{sec:rotation}

The mass and radius of a planet informs us on its global
composition. Because planets are also rotating, one is allowed to
obtain more information on their deep interior structure. 
The hydrostatic equation becomes more complex however:
\begin{equation}
{\bfnab P\over \rho}=\bfnab\left(G\int\!\!\!\int\!\!\!\int 
{\rho(\bfr')\over |\bfr - \bfr'|}d^3\bfr'\right) - \bfOm\times(\bfOm\times\bfr),
\label{eq:full_hydrostat}
\end{equation}
where $\bfOm$ is the rotation vector. 
The resolution of eq.~(\ref{eq:full_hydrostat}) is a complex
problem. It can however be somewhat simplified by assuming that
$|\bfOm|\equiv\omega$ is such that the centrifugal force can be
derived from a potential. The hydrostatic equilibrium then writes
$\nabla P = \rho \nabla U$, and the {\it figure} of the rotating
planet is then defined by the $U=\rm cte$ level surface. 

One can show (e.g. Zharkov \& Trubitsyn 1978) that the hydrostatic
equation of a fluid planet can then be written in terms of the mean
radius $\rbar$ (the radius of a sphere containing the same volume as
that enclosed by the considered equipotential surface):
\begin{equation}
{1\over \rho}\dpar{P}{\rbar}=-{Gm\over \rbar^2}+{2\over 3}\omega^2
\rbar + {GM\over \bar{R}^3} \rbar\varphi_\omega,
\end{equation}
where $M$ and $\bar{R}$ are the total mass and mean radius of the
planet, and $\varphi_\omega$ is a slowly varying function of
$\rbar$. (In the case of Jupiter, $\varphi_\omega$ varies from about
$2\times 10^{-3}$ at the center to $4\times 10^{-3}$ at the surface.)
Equations~(\ref{eq:dtdr}-\ref{eq:dldr}) remain the same with the
hypothesis that the level surfaces for the pressure, temperature, and
luminosity are equipotentials.  The significance of rotation is
measured by the ratio of the centrifugal acceleration to the gravity:
\begin{equation}
q={\omega^2 \req^3\over GM}.
%;\qquad \qbar={\omega^2 \bar{R}^3\over GM}.
\end{equation}

The external gravitational potential of the planet is (assuming
hydrostatic equilibrium):
\begin{equation}
V_{\rm ext}(r,\cos\theta)={GM\over r}\left[1-\sum_{n=1}^\infty\left(a\over
r\right)^{2n}J_{2n}P_{2n}(\cos\theta)\right],
\end{equation}
where the coefficients $J_{2n}$ are the planet's {\it gravitational
moments}, and the $P_{2n}$ are Legendre polynomials. The $J$'s can be
measured by a spacecraft coming close to the planet, preferably on a
polar orbit. Together with the mass, this provides a constraint on the
interior density profile (see Zharkov \& Trubitsyn 1974):
\begin{eqnarray*}
M&=&\int\!\!\!\int\!\!\!\int \rho(r,\theta) d^3\tau, \\
J_{2i} &=& -{1\over M R_{\rm eq}^{2i}}\int\!\!\!\int\!\!\!\int \rho(r,\theta) r^{2i}
P_{2i}(\cos\theta) d^3\tau,
\end{eqnarray*}
where $d\tau$ is a volume element and the integrals are performed over
the entire volume of the planet.

\begin{figure}[htbp]
\resizebox{10cm}{!}{\includegraphics[angle=0]{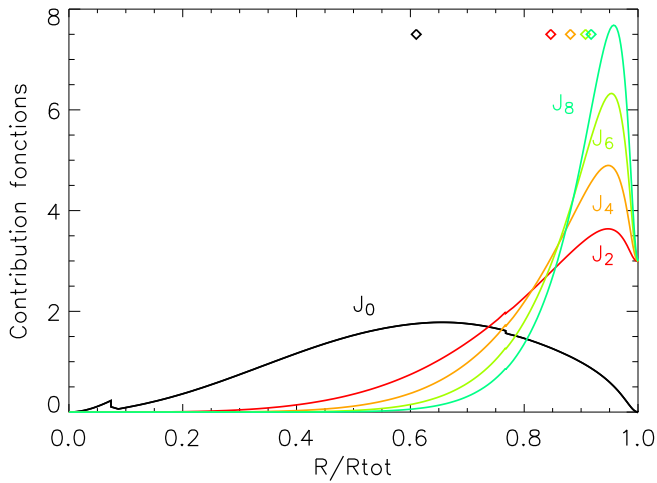}}
%\centerline{\hbox{\psfig{file=poly_index.ps,width=10cm}}}
\caption{Contribution of the level radii to the gravitational moments
  of Jupiter. $J_0$ is equivalent to the planet's mass. The small
  discontinuities are caused by the following transitions, from left
  to right: core/envelope, helium rich/helium poor
  (metallic/molecular). Diamonds indicate the median radius for each
  moment.}
\label{fig:contrib}
\end{figure}

Figure~\ref{fig:contrib} shows how the different layers inside a
planet contribute to the mass and the gravitational moments. The
figure applies to Jupiter, but would remain very similar for other
planets. Measured gravitational moments thus provide information on
the external levels of a planet. It is only indirectly, through the
constraints on the outer envelope that the presence of a central core
can be infered. As a consequence, it is impossible to determine this
core's state (liquid or solid), structure (differentiated, partially
mixed with the envelope) and composition (rock, ice, helium...).

For planets outside the solar system, although measuring their
gravitational potential is utopic, their oblateness may be reachable
with future space transit observations (Seager \& Hui
2002). Since the oblateness $e$ is, to first order, proportionnal to
$q$:
\begin{equation}
e={\req\over\req-\rpol}\approx \left({3\over 2}\Lambda_2+{1\over 2}\right)q
\end{equation}
(where $\Lambda_2=J_2/q\approx 0.1$ to 0.2), it may be possible to
obtain their rotation rate, or with a rotation measured from another
method, a first constraint on their interior structure.

\section{Jupiter, Saturn, Uranus and Neptune}

\subsection{Main observational data}

The mass of the giant planets can be obtained with great accuracy from
the observation of the motions of their natural satellites: 317.834,
95.161, 14.538 and 17.148 times the mass of the Earth ($1\mea =
5.97369\times 10^{27}\g$) for Jupiter, Saturn, Uranus and Neptune,
respectively. The more precise determination of their gravity fields
listed in table~\ref{tab:moments} have been obtained by the {\it
  Pioneer} and {\it Voyager} space missions.

\begin{table}[htb]
\begin{center}
\caption{Characteristics of the gravity fields and radii}
\label{tab:moments}
\small
\begin{tabular}{l r@{.}l r@{.}l r@{.}l r@{.}l} \hline\hline 
	&\multicolumn{2}{c}{\bf Jupiter} &\multicolumn{2}{c}{\bf
Saturn} &\multicolumn{2}{c}{\bf Uranus} &\multicolumn{2}{c}{\bf
Neptune} \\ \hline
$M\p{-29}$ [g] & 18&986112(15)\ti{a} & 5&684640(30)\ti{b}
 & 0&8683205(34)\ti{c} & 1&0243542(31)\ti{d} \\
$R_{\rm eq}\p{-9}$ [cm] &  7&1492(4)\ti{e} & 6&0268(4)\ti{f} 
 & 2&5559(4)\ti{g} & 2&4766(15)\ti{g} \\
$R_{\rm pol}\p{-9}$ [cm] &  6&6854(10)\ti{e} & 5&4364(10)\ti{f} 
 & 2&4973(20)\ti{g} & 2&4342(30)\ti{g} \\
$\bar{R}\p{-9}$ [cm] & 6&9894(6)\ti{h} & 5&8210(6)\ti{h} & 2&5364(10)\ti{i} &
  2&4625(20)\ti{i} \\
$\bar{\rho}$ [$\gcc$] & 1&3275(4) & 0&6880(2) & 1&2704(15) & 1&6377(40) \\
$J_2\p{2}$ & 1&4697(1)\ti{a} & 1&6332(10)\ti{b}
 & 0&35160(32)\ti{c} & 0&3539(10)\ti{d}  \\
$J_4\p{4}$ & $-5$&84(5)\ti{a} & $-9$&19(40)\ti{b}
 & $-0$&354(41)\ti{c} & $-0$&28(22)\ti{d} \\
$J_6\p{4}$ & 0&31(20)\ti{a} & 1&04(50)\ti{b}
 & \multicolumn{2}{c}{\dots} & \multicolumn{2}{c}{\dots} \\
$P_{\omega}\p{-4}$ [s] & 3&57297(41)\ti{j} & 3&83577(47)\ti{j}
 & 6&206(4)\ti{k} & 5&800(20)\ti{l} \\
$q$ & 0&08923(5) & 0&15491(10) & 0&02951(5) & 0&02609(23) \\
$C/M\req^2$ & 0&258 & 0&220 & 0&230 & 0&241 \\

\hline\hline
\multicolumn{9}{p{11.2cm}}{%
The numbers in parentheses are the uncertainty in the last digits 
of the given value. The value of the gravitational constant used to
calculate the masses of Jupiter and Saturn is $G=6.67259\p{-8}\,\rm
dyn.cm^2.g^{-1}$ (Cohen \& Taylor, 1987).}\\
\multicolumn{9}{l}{\ti{a} Campbell \& Synott (1985)}\\
\multicolumn{9}{l}{\ti{b} Campbell \& Anderson (1989)}\\
\multicolumn{9}{l}{\ti{c} Anderson \etal\ (1987)}\\
\multicolumn{9}{l}{\ti{d} Tyler \etal\ (1989)}\\
\multicolumn{9}{l}{\ti{e} Lindal \etal\ (1981)}\\
\multicolumn{9}{l}{\ti{f} Lindal \etal\ (1985)}\\
\multicolumn{9}{l}{\ti{g} Lindal (1992)}\\
\multicolumn{9}{l}{\ti{h} From 4th order figure theory}\\
\multicolumn{9}{l}{\ti{i} $(2\req+\rpol)/3$ (Clairaut's approximation)}\\
\multicolumn{9}{l}{\ti{j} Davies \etal\ (1986)}\\
\multicolumn{9}{l}{\ti{k} Warwick \etal\ (1986)}\\
\multicolumn{9}{l}{\ti{l} Warwick \etal\ (1989)}\\
\end{tabular}
\normalsize
\end{center}
\end{table}

Table~\ref{tab:moments} also indicates the radii obtained with the
greatest accuracy by radio-occultation experiments. By convention,
these radii and gravitational moments correspond to the 1\,bar
pressure level. The rotation periods are measured from the variations
of the planets' magnetic fields (system III) and are believed to be
tied to the interior rotation. The giant planets are relatively fast
rotators, with periods of about 10 hours for Jupiter and Saturn,
and about 17 hours for Uranus and Neptune. The fact that this fast
rotation visibly affects the figure (shape) of these planets is seen
by the significant difference between the polar and equatorial radii.

A first result obtained from the masses and radii (using the planets'
{\it mean} radii, as defined in section~\ref{sec:rotation}) indicated
in Table~\ref{tab:moments} is the fact that these planets have low
densities. These densities are similar, but considering that
compression strongly increases with mass, one is led to a
sub-classification between the hydrogen-helium giant planets Jupiter
and Saturn, and the ``ice giants'' Uranus and Neptune.

The values of the axial moment of inertia $C$ have been calculated
using the Radau-Darwin approximation (Zharkov \& Trubitsyn 1978). Our
four giant planets all have an axial moment of inertia substantially
lower than the value for a sphere of uniform density, \ie\
$2/5\,MR^2$, indicating that they have dense central regions. This
does not necessarily mean that they possess a core, but simply that
the density profile departs significantly from a uniform value. 

\begin{table}[htb]
\begin{center}
\caption{Energy balance as determined from Voyager IRIS data\ti{a}.}
\label{tab:flux}
\small
\begin{tabular}{l r@{.}l r@{.}l r@{.}l r@{.}l} \hline \hline
	& \multicolumn{2}{c}{\bf Jupiter} &
\multicolumn{2}{c}{\bf Saturn} &\multicolumn{2}{c}{\bf Uranus} 
&\multicolumn{2}{c}{\bf Neptune}  \\
\hline 
Absorbed power [$10^{23}$\, erg.s$^{-1}$] & 50&14(248) &
11&14(50) & 0&526(37)& 0&204(19)\\
Emitted power [$10^{23}$\, erg.s$^{-1}$] & 83&65(84)&
19&77(32)& 0&560(11)& 0&534(29)\\
Intrinsic power [$10^{23}$\, erg.s$^{-1}$]&
33&5(26)& 8&63(60)& 0&034(38) 
%multicolumn{2}{c}{0.034\,\parbox{2em}{\tiny $+0.038$ $-0.034$}} 
& 0&330(35)\\
%L/M$\,^{\rm (c)}$ [$10^7$\, erg.g$^{-1}$.s$^{-1}$] & $17.6& 1.4$ &
%$15.2& 1.1$ & 0.392\,\parbox{2em}{\tiny $+0.441$ $-0.392$} & 
%$3.22& 0.34$ \\
Intrinsic flux [erg.s$^{-1}$.cm$^{-2}$] & 5440&(430)& 2010&(140)& 
%multicolumn{2}{c}{42\,\parbox{1.2em}{\tiny +47$ $-42$}} & 433& 46 \\
42&(47) & 433&(46) \\
Bond albedo [] & 0&343(32)& 0&342(30)& 0&300(49)& 0&290(67)\\
Effective temperature [K] & 124&4(3)& 95&0(4)& 59&1(3)& 59&3(8)\\ 
1-bar temperature\ti{b} [K] & 165&(5) & 135&(5) & 76&(2) & 72&(2) \\
\hline\hline 
\multicolumn{9}{l}{\ti{a} After Pearl \& Conrath (1991)} \\
\multicolumn{9}{l}{\ti{b} Lindal (1992)} 
\end{tabular}
\normalsize
\end{center}
\end{table}

Jupiter, Saturn and Neptune are observed to emit significantly more
energy than they receive from the Sun (see
Table~\ref{tab:flux}). The case of Uranus is less clear. Its
intrinsic heat flux $F_{\rm int}$ is significantly smaller than that
of the other giant planets. Detailed modeling of its atmosphere
however indicate that $F_{\rm int}\wig{>}60\rm\,erg\,cm^{-2}\,s^{-1}$
(Marley \& McKay 1999). 
With this caveat, all four giant planets can be said to emit more
energy than they receive from the Sun. 
Hubbard (1968) showed in the case of Jupiter that this can be
explained simply by the progressive contraction and cooling of the
planets. 

It should be noted that the 1 bar temperatures listed in
table~\ref{tab:flux} are retrieved from
radio-occultation measurements using a helium to hydrogen ratio which,
at least in the case of Jupiter and Saturn, was shown to be
incorrect. The new values of $Y$ are found to lead to increased
temperatures by $\sim 5\K$ in Jupiter and $\sim 10\K$ in Saturn (see
Guillot 1999a). However, the Galileo probe found
a 1 bar temperature of $166\K$ (Seiff
\etal\ 1998), and generally a good agreement with the
Voyager radio-occultation profile with the wrong He/H$_2$ value.

\subsection{Atmospheric composition}

The most important components of the atmospheres of our giant planets
are also among the most difficult to detect: H$_2$ and He have a zero
dipolar moment. Also their rotational lines are either weak or
broad. On the other hand, lines due to electronic transitions
correspond to very high altitudes in the atmosphere, and bear little
information on the structure of the deeper levels. 
The only robust result concerning the abundance of helium in a
giant planet is by {\it in situ} measurement by the Galileo probe in
the atmosphere of Jupiter (von Zahn \etal\ 1998). The helium mole
fraction (\ie\ number of 
helium atoms over the total number of species in a given volume) is
$q_{\rm He}=0.1359\pm 0.0027$. The helium mass mixing ratio $Y$ (\ie\
mass of helium atoms over total mass) is constrained by its ratio over
hydrogen, $X$: $Y/(X+Y)=0.238\pm 0.05$. This ratio is by coincidence
that found in the Sun's atmosphere, but because of helium
sedimentation in the Sun's radiative zone, it was larger in the
protosolar nebula: $Y_{\rm proto}=0.275\pm 0.01$ and $(X+Y)_{\rm
proto}\approx 0.98$. Less helium is therefore found in
the atmosphere of Jupiter than inferred to be present when the planet
formed. 

Helium is also found to be depleted compared to the protosolar value
in Saturn's atmosphere. However, in this case the analysis is
complicated by the fact that Voyager radio occultations apparently led
to a wrong value. The current adopted value is now $Y=0.18-0.25$
(Conrath \& Gautier 2000), in agreement with values predicted by
interior and evolution models (Guillot 1999a; Hubbard \etal\ 1999). 
Finally, Uranus and Neptune are found to have near-protosolar helium
mixing ratios, but with considerable uncertainty (Gautier \& Owen
1989).

\begin{figure}[htbp]
\resizebox{12cm}{!}{\includegraphics[angle=0]{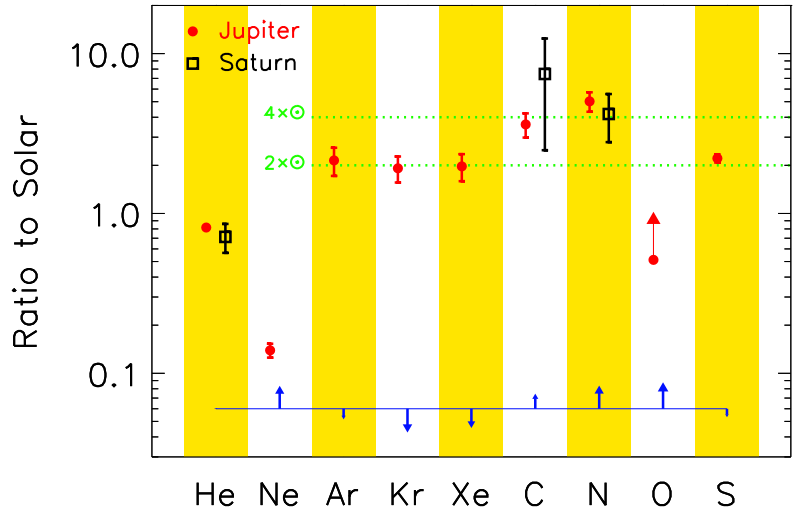}}
\caption{Elemental abundances measured in the tropospheres of Jupiter
  (circles) and Saturn (squares) in units of their abundances in the
  protosolar nebula. The elemental abundances for Jupiter are derived
  from the in situ measurements of the Galileo probe (e.g. Mahaffy et
  al. 2000; Atreya et al. 2003). Note that the oxygen abundance is
  considered to be a minimum value due to meteorological effects
  (Roos-Serote et al. 2004). The abundances for Saturn are
  spectroscopic determination (Atreya et al. 2003 and references
  therein). The solar or protosolar abundances used as a reference are
  from Lodders (2003). The arrows show how abundances are affected by
  changing the reference protosolar abundances from those of Anders \&
  Grevesse (1989) to those of Lodders (2003). The horizontal dotted
  lines indicate the locus of a uniform 2- and 4-times solar
  enrichment in all elements except helium and neon, respectively.}
\label{fig:enrichments}
\end{figure}

The abundance of ``heavy elements'', i.e. elements other than hydrogen
and helium, bears crucial information for the understanding of the
processes that led to the formation of these planets. Again, the most
precise measurements are for Jupiter, thanks to the Galileo probe.  As
shown by \rfig{enrichments}, most of the heavy elements are enriched
by a factor 2 to 4 compared to the solar abundance (Niemann \etal\
1998; Owen \etal\ 1999). One exception is neon, but an explanation is
its capture by the falling helium droplets (Roustlon \& Stevenson
1995). Another exception is water, but this molecule is affected by
meteorological processes, and the probe was shown to have fallen into
a dry region of Jupiter's atmosphere. There are strong indications
that its abundance is at least solar. Possible very high interior
abundances ($\sim 10$ times the solar value) have also been suggested
%, either to explain waves
%propagation after the Shoemaker-Levy 9 impacts (Ingersoll \etal\ 1994)
%or 
as a scenario to explain the delivery of heavy elements to the
planet (Gautier \etal\ 2001, Hersant et al. 2004).

In the case of Saturn, both carbon in the form of methane and nitrogen
as ammonia appear to be significantly enriched, but with large error
bars (Atreya et al. 2003). In Uranus and Neptune, methane is
probably between 30 and 60 times the solar value (Gautier \&
Owen 1989; Hersant et al. 2004).

\subsection{Interior models: Jupiter and Saturn}

\begin{figure}[htbp]
\centerline{\resizebox{12cm}{!}{\includegraphics[angle=90,bb=4.5cm 7cm 15cm 24cm]{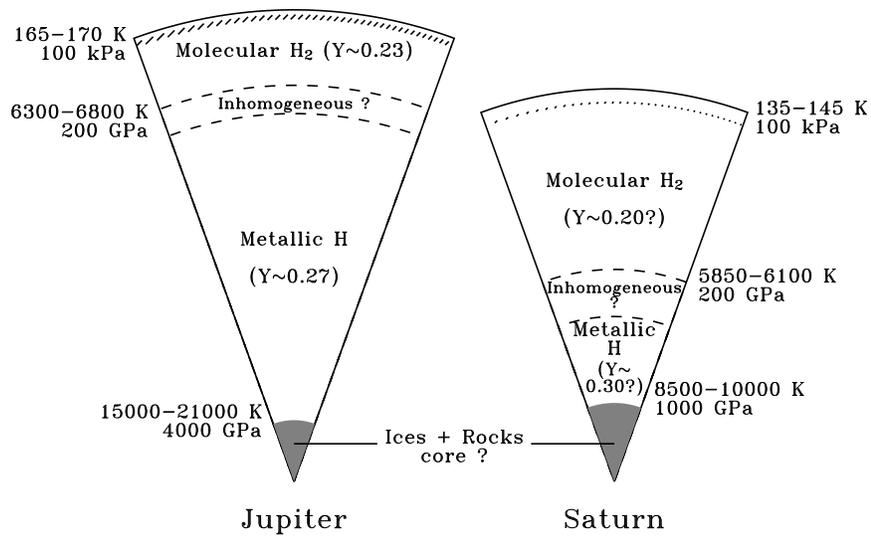}}}
\caption{Schematic representation of the interiors of Jupiter and
  Saturn. The range of temperatures is estimated using homogeneous
  models and including a possible radiative zone indicated by the hashed
  regions. Helium mass mixing ratios $Y$ are indicated. The size of the
  central rock and ice cores of Jupiter and Saturn is very uncertain
  (see text). In the case of Saturn, the inhomogeneous region may
  extend down all the way to the core which would imply the formation
  of a helium core. [Adapted from Guillot 1999b].}
\label{fig:intjupsat}
\end{figure}

As illustrated by fig.~\ref{fig:intjupsat}, the simplest interior
models of Jupiter and Saturn matching all observational constraints
assume the presence of three main layers: (i) an outer hydrogen-helium
envelope, whose global composition is that of the deep atmosphere;
(ii) an inner hydrogen-helium envelope, enriched in helium because the
whole planet has to fit the H/He protosolar value; (iii) a central
dense core. Because the planets are believed to be mostly convective,
these regions are expected to be globally homogeneous. (Many
interesting thermochemical transformations take place in the deep
atmosphere, but they are of little concern to
us). 

A large part of the uncertainty in the models lies in the existence
and location of an inhomogeneous region in which helium separates from
hydrogen to form helium-rich droplets that fall deeper into the planet 
due to their larger density. Models have generally assumed this region
to be relatively narrow, because helium was thought to be most
insoluble in low-pressure metallic hydrogen (e.g. Stevenson
1982). However, DFT calculations have indicated that the critical
temperature for helium demixing may rise with pressure (Pfaffenzeller
et al. 1995), presumably in the regime where hydrogen is only partially
ionized and bound states remain. This opens up the possibility that the
inhomogeneous regions may be more extended. In particular, in the case
of Saturn, Fortney \& Hubbard (2003) have shown that explaining
Saturn's age may require that helium fall all the way to the core,
thereby yielding the formation of a helium core (or of a helium shell
around a rock or ice core). 

\begin{figure}[htbp]
\resizebox{11cm}{!}{\includegraphics[angle=0]{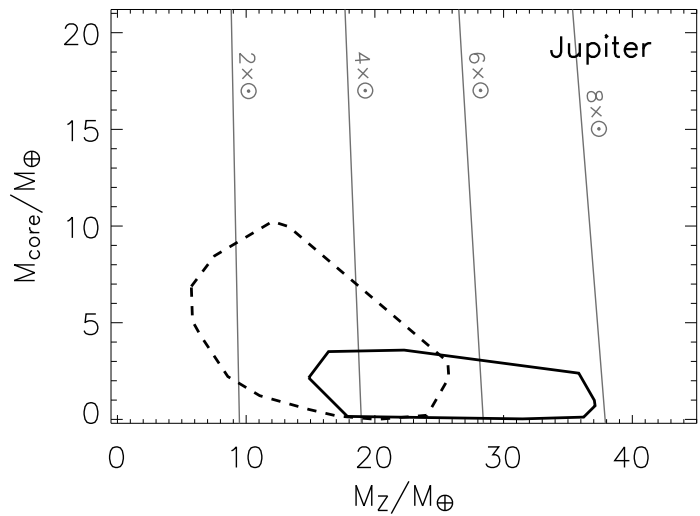}}
\caption{Constraints on Jupiter's interior structure based on Saumon
  \& Guillot (2004). The value of the core mass ($M_{\rm core}$) is
  shown in function of the mass of heavy elements in the envelope
  ($M_{Z}$) for models matching all available observational
  constraints. The dashed region corresponds to models matching the
  laser compression experiments. The plain box corresponds to models
  matching the pulsed power and convergent shock compression
  experiments (see text). Grey lines indicate the values of $M_Z$ that
  imply uniform enrichments of the envelope in heavy elements by
  factors 2 to 8 times the solar value ($Z_\odot=0.0149$),
  respectively.}
\label{fig:zel_jup}
\end{figure}

With these caveats, the three-layer models can be used as a useful
guidance to a necessarily hypothetical ensemble of allowed structures
and compositions of Jupiter and Saturn.  Figure~\ref{fig:zel_jup}
shows such an ensemble for Jupiter, based on calculations by Saumon \&
Guillot (2004). The calculations assume that only helium is
inhomogeneous in the envelope (the abundance of heavy elements is
supposed to be uniform accross the molecular/metallic hydrogen
transition). Many sources of uncertainties are included however; among
them, the most significant are on the equations of state of hydrogen
and helium, the uncertain values of $J_4$ and $J_6$, the presence of
differential rotation deep inside the planet, the location of the
helium-poor to helium-rich region, and the uncertain helium to
hydrogen protosolar ratio.

These results show that Jupiter's core is smaller than $\sim 10\mea$,
and that its global composition is pretty much unknown (between 10 to
42$\mea$ of heavy elements in total). The models indicate that Jupiter
is enriched compared to the solar value, particularly with the new,
low value of $Z_\odot$ (Lodders 2003) used in
fig.~\ref{fig:zel_jup}. This enrichment could be compatible with a
global uniform enrichment of all species near the atmospheric Galileo
values. Alternatively, species like oxygen (as mostly water) may be
significantly enriched.

Most of the constraints are derived from the values of the radius (or
equivalently mass) and of $J_2$. The measurement of $J_4$ allows to
further narrow the ensemble of possible models, and in some cases, to
rule out EOS solutions (in particular those indicating relatively
large core masses, between 10 and 20$\mea$).  As discussed in Guillot
(1999a) and Saumon \& Guillot (2004), most of the uncertainty in the
solution arises because very different hydrogen EOSs are possible. The
fact that more laboratory and numerical experiments seem to indicate
relatively low-compressions for hydrogen at Mbar pressures points
towards smaller core masses and a larger amount of heavy elements in
the planet (plain box in fig.~\ref{fig:zel_jup}). However, this relies
on uncertain temperature gradients, because the EOSs are based on
laboratory data obtained at temperatures higher than those relevant to
the planetary interiors.

Results slightly outside the boxes of fig.~\ref{fig:zel_jup} are
possible in the presence of a discontinuity of the abundance of heavy
elements in the interior. Thus, Guillot (1999a) found slightly larger
core masses (up to $12\mea$) in the case of the Saumon-Chabrier EOS
with a first order plasma-phase transition.

\begin{figure}[htbp]
\resizebox{11cm}{!}{\includegraphics[angle=0]{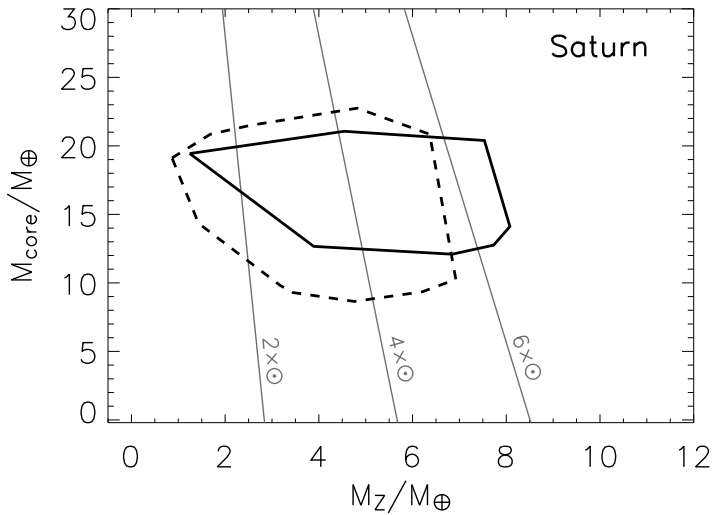}}
\caption{Same as fig.~\ref{fig:zel_jup} in the case of Saturn. Note
  that smaller core masses could result either from allowing a
  variation of the abundance of heavy near the molecular/metallic
  transition (Guillot 1999a), or from the presence of a helium shell
  around the core (Fortney \& Hubbard 2003).}
\label{fig:zel_sat}
\end{figure}

In the case of Saturn (\rfig{zel_sat}), the solutions depend less on
the hydrogen EOS because the Mbar pressure region is comparatively
smaller. The total amount of heavy elements present in the planet can
therefore be estimated with a better accuracy than for
Jupiter. However, because Saturn's metallic region is deeper into the
planet, it mimics the effect that a central core would have on
$J_2$. If we allow for variations in the abundance of heavy elements
together with the helium discontinuity, then the core mass can become
much smaller, and even solutions with no core can be found (Guillot
1999a). These solutions depend on the hypothetic phase separation of
an abundant species (e.g. water), and generally cause an energy
problem because of the release of considerable gravitational energy. 
However, another possibility is through the formation of an almost
pure helium shell around the central core, which could lower
the core masses by up to $~7\mea$ (Fortney \& Hubbard 2003; Hubbard,
personnal communication).

\subsection{Interior models: Uranus and Neptune}

\begin{figure}[htbp]
\centerline{\resizebox{11cm}{!}{\includegraphics[angle=90,bb=4.5cm 4.5cm 15cm 24cm]{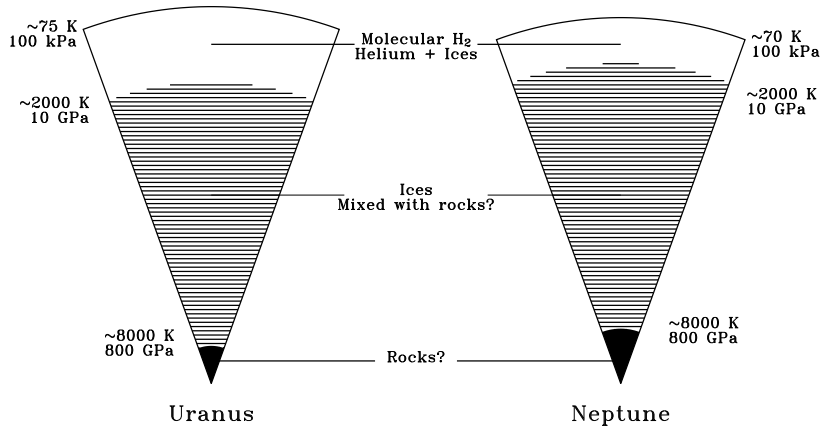}}}
%\vspace*{-1cm}
\caption{Schematic representation of the interiors of Uranus and
  Neptune. [Adapted from Guillot 1999b].}
\label{fig:inturanep}
\end{figure}

Although the two planets are relatively similar, \rfig{m-r}\ already
shows that Neptune's larger mean density compared to Uranus has to be
due to a slightly different composition: either more heavy elements
compared to hydrogen and helium, or a larger rock/ice ratio.  The
gravitational moments impose that the density profiles lie close to
that of ``ices'' (a mixture initially composed of H$_2$O, CH$_4$ and
NH$_3$, but which rapidly becomes a ionic fluid of uncertain chemical
composition in the planetary interior), except in the outermost
layers, which have a density closer to that of hydrogen and helium
(Marley \etal\ 1995; Podolak \etal\ 2000). As illustrated in
\rfig{inturanep}, three-layer models of Uranus and Neptune consisting
of a central ``rocks'' core (magnesium-silicate and iron material), an
ice layer and a hydrogen-helium gas envelope have been calculated
(Podolak \etal\ 1991; Hubbard \etal\ 1995).

The fact that models of Uranus assuming homogeneity of each layer and
adiabatic temperature profiles fail in reproducing its gravitational
moments seem to imply that substantial parts of the planetary interior
are not homogeneously mixed (Podolak \etal\ 1995).  This could explain
the fact that Uranus' heat flux is so small: its heat would not be
allowed to escape to space by convection, but through a much slower
diffusive process in the regions of high molecular weight
gradient. Such regions would also be present in Neptune, but much
deeper, thus allowing more heat to be transported outward.  The
existence of these non-homogeneous, partially mixed regions are
further confirmed by the fact that if hydrogen is supposed to be
confined solely to the hydrogen-helium envelope, models predict
ice/rock ratios of the order of 10 or more, much larger than the
protosolar value of $\sim\,$2.5. On the other hand, if we impose the
constraint that the ice/rock ratio is protosolar, the overall
composition of both Uranus and Neptune is, by mass, about 25\% rocks,
$60-70$\% ices, and $5-15$\% hydrogen and helium (Podolak \etal\ 1991,
1995; Hubbard \etal\ 1995). Assuming both ices and rocks are present
in the envelope, an upper limit to the amount of hydrogen and helium
present is $\sim 4.2\mea$ for Uranus and $\sim 3.2\mea$ for Neptune
(Podolak \etal\ 2000). A lower limit of $\sim 0.5\mea$ for both
planets can be inferred by assuming that hydrogen and helium are only
present in the outer envelope at $P\wig<100$\,kbar.

\subsection{{\it Are the interiors adiabatic?}}

As discussed, the near-adiabaticity of the interiors of the giant
planets is a consequence of the rapid rise of opacities with
increasing pressure and temperatures. Several exceptions are possible:

(i) In the ``meteorological layer'', the temperature gradient could
become either subadiabatic (because of latent heat release and moist
convection) or superadiabatic (because of molecular weight gradients
created by condensation and precipitation). Locally, a depletion of an
efficient radiative absorber (e.g. water or methane) could imply that
convection is suppressed, either because of a lowered radiative
gradient, or because sunlight can then be deposited to this level.  In
Uranus and Neptune, a superadiabatic region at $P\sim 1-2\,$bar is
correlated with methane condensation (Lindal 1992, Guillot 1995). In
Jupiter, the Galileo probe measured a nearly-adiabatic profile, with a
slight static stability ($N<0.2\rm\,K\,km^{-1}$) down to 20\,bars
(Magalh\~aes et al. 2002).

(ii) At the Plasma Phase Transition between molecular and metallic
hydrogen, if it exists, with an entropy jump that could be of order
$1\,k_{\rm B}$/baryon (Stevenson \& Salpeter 1977; Saumon et
al. 1995).

(iii) In the hydrogen-helium phase separation region, where a slow
droplet formation may inhibit convection and yield a significant
superadiabacity (Stevenson \& Salpeter 1977).

(iv) Near the core/envelope interface (whether it is abrupt or not)
where an inhibiting molecular weight gradient occurs and, in the case
of Jupiter, conduction might play a role.

(v) Throughout the planets, even though mixing-length arguments
predict that the superadiabacity is extremely small ($\sim 10^{-6}$ or
less), rotation and magnetic fields may increase it, although probably
by modest amounts (Stevenson 1982; See also discussion in Guillot et
al. 2004).

\subsection{{\it What are the ages of our giant planets?}}

If we understand something of the formation of our Solar System and of
other stars, our giant planets should have formed 4.55\,Ga ago
(e.g. Bodenheimer \& Lin 2002). The {\it model} ages show significant
deviations from that value, however. 

In the case of Jupiter, the present radius and luminosity are obtained
after 3.5 to 5.5\,Ga of evolution, but most realistic EOSs predict
ages above 4.5\,Ga (Saumon \& Guillot 2004). Several processes, among
which core erosion, could lead to a reduction of that value (Guillot
et al. 2004).  For Saturn, homogeneous evolution models predict ages
of order 2\,Ga (Stevenson 1982; Saumon et al. 1992; Guillot et
al. 1995). In both planets, the presence of a phase separation of
helium is likely and would tend to lengthen the cooling.

The case of Uranus and Neptune is less clear-cut because of the
uncertainties both on the properties of their atmospheres (in
particular their evolution with time), and on the global specific heat
of material inside. It appears however that both planets have
luminosities that are too small. This could be due to a cold start
(relatively low initial temperatures), a rapid loss of the internal
heat, or a strong molecular weight gradient that prevents interior
regions from cooling (Podolak et al. 1991; Hubbard et al. 1995).

\subsection{{\it Do some elements separate from hydrogen at high
    pressures? Where?}} 

Helium is strongly suspected of separating from hydrogen in Jupiter
and Saturn because its lower than protosolar abundance in the
atmosphere, and in Saturn because without this additional energy
source, the planet would evolve to its present state in $\sim
2\,$Ga. However, it has not been shown so far that a hydrogen helium
mixture at Mbar pressures has a critical demixing temperature that is
above that required in Jupiter and Saturn.

Helium demixing should occur in the metallic hydrogen region, but it
is not clear that the critical temperature should decrease with
pressure as for fully ionized plasmas (Stevenson 1982), or increase
with pressure (Pfaffenzeller et al. 1995). The first scenario would imply the
existence of a small inhomogeneous region near the molecular/metallic
transition as illustrated in \rfig{intjupsat}. The second one would
yield a more extended inhomogeneous region.

Evolution models including the two phase diagrams by Fortney \&
Hubbard (2003) show that in order to reconcile Saturn's age with that
of the Solar System and the atmospheric helium abundance derived by
Conrath \& Gautier (2000), sufficient energy $\sim \Delta M_{\rm
He}gH$ is required. This implies maximizing $H$, the distance of
sedimentation of helium droplets, and hence favors the
Pfaffenzeller-type phase diagram and the formation of a helium core. 

The question of a phase separation of other elements is still open. It
is generally regarded as unlikely at least in Jupiter and Saturn
because of their small abudances relative to hydrogen and the fact
that the critical demixing temperature depends exponentially on that
abundance.

\subsection{{\it How do the planetary interiors rotate?}}

Interior rotation is important because it affects the gravitational
moments and their interpretation in terms of density profiles (Zharkov
\& Trubitsyn 1978). It is presently not known whether the observed
atmospheric zonal flow patterns are tied to the planetary interiors or
whether they are surface phenomena, with the interior rotating close
to a solid body with the rate given by the magnetic field. Interior
rotation affects more significantly gravitational moments of higher
order. Using extrema set by solid rotation and by a model in which the
zonal wind pattern is projected into a cylindrical rotation (Hubbard
1982), one can show that interior rotation introduces an uncertainty
equivalent to the present error bar for $J_4$, of the order of the
spread in interior models for $J_6$, and that becomes dominant for
$J_8$ and above. Measurements of high order gravitational moments
$J_8-J_{14}$ should tell whether atmospheric zonal flow penetrate into
the deep interior or whether the deep rotation is mainly solid
(Hubbard 1999).

\subsection{{\it What can we tell of the giant planets' cores? Are
    they primordial?}}

Confronted to diagrams such as figs.~\ref{fig:intjupsat} and
\ref{fig:inturanep}, there is the tendency to think that the giant
planets cores as well defined, separate entities. It is not
necessarily the case: first, as shown by \rfig{contrib}, solutions
with a well-defined central core are equivalent to solutions with
cores that have been diluted into the central half of the
planet. Second, convection does not necessarily guarantee the presence
of globally homogeneous regions, and can efficiently oppose the
settling of species, as observed in thermohaline convection. Finally,
the history of core formation, and in particular the epoch at which
planetesimals were accreted and their sizes matter (e.g. Stevenson
1985).

Once formed, the cores of the giant planets are difficult to erode, as
this demands both that heavy elements are (at least partially) soluble
in the hydrogen helium envelope, and that enough energy is present to
overcome the molecular weight barrier that is created (Stevenson
1982). However, in the case of Jupiter at least, the second condition
may not be that difficult to obtain, as only 10\% of the energy in the
first convective cell (in the sense of the mixing length approach)
needs to be used to dredge up about $20\mea$ of core material (Guillot
et al. 2004). Evaluating whether the first condition is satisfied
would require knowing the core's composition and its state, but one
can nevertherless note that the initially high central temperatures
($\sim 30,000\,$K) favor solubility. Such an efficient erosion would
not occur in Saturn (and much less so in Uranus and Neptune) because
of its smaller total mass.

\subsection{{\it Do we understand the planets' global compositions?}}

This may be the hardest question because it requires tying all the
different aspects of planet formation to the observations of the
atmospheres of the giant planets and the constraints on their interior
structures. So far, most of the focus has been on explaining the
presence of a central core of $\sim 10\mea$ in Jupiter, Saturn, Uranus
and Neptune. The new interior data suggest that Jupiter's core is
probably smaller, and that Saturn's may be larger. More importantly,
the envelopes of all planets appear to be enriched in heavy elements,
and this has to be explained as well. 

The possibility that Jupiter could have been formed by a direct
gravitational instability (e.g. Boss 2000) may be appealing in view of
its small inferred core. However, the enrichment of its envelope in
heavy elements is difficult to explain within that scenario, given the
low accretion rate of a fully-formed Jupiter (Guillot \& Gladman
2000). 

The leading scenario therefore remains the standard ``core accretion''
scenario (Pollack et al. 1996), with the addition that Jupiter,
Saturn, Uranus and Neptune were closer together (5-20\,AU) just after
their formation (Levison \& Morbidelli 2003). While
this scenario require core masses $\wig{>}10\mea$, the possibility of
an erosion of Jupiter's core is appealing because it would both
explain the difference in size with Saturn and an enrichment of its
envelope. Although more limited, a small $\sim 2\mea$ erosion of
Saturn's core could provide part of the enrichment of the envelope
(Guillot et al. 2004).

The fact that Jupiter's atmosphere is also enriched in noble gases, in
particular Ar which condenses at very low temperatures ($\sim 30\K$)
is still a puzzle. Presently invoked explanations include a clathration
of noble gases in ices (Gautier et al. 2001; Hersant et al. 2004), and
the delivery of planetesimals formed at very low temperatures (Owen
et al. 1999).

Finally, the large enrichments in C and possibly N of the atmospheres
of Uranus and Neptune probably indicate that a significant mass of
planetesimals ($\wig> 0.1\mea$) impacted the planets after they had
captured most of their present hydrogen-helium envelopes. Along with
the other problems related to this section, this requires quantitative
work.  

%\subsection{{\it How are the magnetic fields generated?}}

\section{Extrasolar planets}

\subsection{Observables}

More than 145 extrasolar planets have been discovered to date (see
J. Schneider's {\it Extrasolar Planets Encyclopedia} on
http://www.obspm.fr/planets), but only those for which a determination
of both the planetary mass and radius are useful for the purposes of
this review. This can only be done for planets which transit in front
of their star, which, by probabilistic arguments limits us to planets
that orbit close to their star. I will therefore only be concerned
with ``Pegasi planets'', giant planets similar to 51 Peg b and
HD209458b (both in the constellation Pegasus), with semi-major
axes smaller than 0.1\,AU. 

\begin{table}
  \caption{Systems with transiting Pegasi planets discovered so far}
  \label{tab:transits}
  \begin{tabular}{lrrrrrr} \hline \hline
 & Age [Ga] & [Fe/H] & a [AU] & $\teq^\star$ [K] & $M_{\rm p}/M_{\rm J}$ &
 $R_{\rm p}/{\rm 10^{10}\,cm}$ \\
{\bf HD209458}\ti{a} & $4-7$ & 0.00(2) & 0.0462(20) & 1460(120) & 0.69(2)
 & 1.02(9) \\
{\bf OGLE-56}\ti{b} & $2-4$ & 0.0(3) & 0.0225(4) & 1990(140) &
 1.45(23) & 0.88(11) \\
 {\bf OGLE-113}\ti{c} & ? & 0.14(14) & 0.0228(6) & 1330(80) &
 0.765(25) & $0.77\!\left(^{+5}_{-4}\right)$ \\
 {\bf OGLE-132}\ti{d} & $0-1.4$ & 0.43(18) & 0.0307(5) & 2110(150)
 & 1.19(13) & 0.81(6) \\
{\bf OGLE-111}\ti{e} & ? & 0.12(28) & 0.0470(10) & 1040(160) &
 0.53(11) & $0.71\!\left(^{+9}_{-4}\right)$ \\
{\bf TrES-1}\ti{f} & ? & 0.00(4) & 0.0393(11) & 1180(140) & 0.75(7) &
 $0.77(4)$\\
    \hline \hline
    \multicolumn{7}{l}{$^\star$ Equilibrium temperature calculated on the basis
 of a zero planetary albedo}\\
    \multicolumn{7}{l}{\ti{a}Cody \& Sasselov (2002), Brown et al. (2001)}\\
    \multicolumn{7}{l}{\ti{b}Torres et al. (2004), Sasselov (2003), Konacki et al. (2003)}\\
    \multicolumn{7}{l}{\ti{c}Bouchy et al. (2004), Konacki et al. (2004)}\\
    \multicolumn{7}{l}{\ti{d}Moutou et al. (2004)}\\
    \multicolumn{7}{l}{\ti{e}Pont et al. (2004)}\\
    \multicolumn{7}{l}{\ti{f}Laughlin et al. (2004), Sozzetti et al. (2004), Alonso et al. (2004)}\\

  \end{tabular}
\end{table}

Six transiting Pegasi planets have been discovered so far. Their main
characteristics are listed in Table~\ref{tab:transits}. The first one,
HD209458b (Charbonneau et al. 2000; Henry et al. 2000), has been shown
to possess sodium in its atmosphere (Charbonneau et al. 2002) and to
have an extended, evaporating atmosphere (Vidal-Madjar et al. 2003,
2004). Four others have been discovered by the photometric OGLE survey
and subsequent radial velocity measurements (Konacki et al. 2003,
2004; Bouchy et al. 2004, Pont et al. 2004). One is a result of the
TrES network survey (Alonso et al. 2004). Present photometric surveys
have a strong detection bias towards very short periods. Associated to
a probability of transiting that is inversely proportional to the
orbital distance, this shows that Table~\ref{tab:transits} represents
only a tiny fraction of planets which may have a low probability of
existence. 

A crucial parameter for the evolution models is the equilibrium
temperature $\teq=T_*\sqrt{R_*/2a}$ (assuming a zero albedo, i.e. that
all incoming stellar light is absorbed by the planetary atmosphere).
With values of $\teq$ between $\sim 1000$ and 2000\,K, the present
sample of transiting planets is already quite rich.

\subsection{Observed vs. calculated radii of ``Pegasi planets''}

Contrary to the giant planets in our Solar System, Pegasi planets are
subject to an irradiation from their central star that is so intense
that the absorbed stellar energy flux is about $\sim 10^4$ times
larger than their intrinsic flux (estimated from \eref{lapprox}, or
calculated directly). The atmosphere is thus prevented from cooling,
with the consequence that a radiative zone develops and governs the
cooling and contraction of the interior (Guillot et
al. 1996). Typically, for a planet like HD209458b, this radiative zone
extends to kbar levels, $T\sim 4000\K$, and is located in the outer
5\% in radius ($0.3\%$ in mass) (Guillot \& Showman 2002).

Problems in the modeling of the evolution of Pegasi planets arise
mostly because of the uncertain outer boundary condition. The intense
stellar flux implies that the atmospheric temperature profile is
extremely dependant upon the opacity sources considered. Depending on
the chosen composition, the opacity data used, the assumed presence of
clouds, the geometry considered, resulting temperatures in the deep
atmosphere can differ by up to $\sim 600\K$ (Seager \& Sasselov 1998,
2000; Goukenleuque et al. 2000; Barman et al. 2001; Sudarsky et
al. 2003; Iro et al. 2004). Because of this problem, and in the
framework of our simple model, the following discussion will be based
on an outer boundary condition at 1\,bar and a fixed temperature
$T_1=1500$ or $2000\K$\footnote{Technically, in order to obtain high
entropy initial conditions I use $T_1\approx \teq(1+L/L_{\rm
eq})^{1/4}$, but the precise form does not matter as long as $L\ll
L_{\rm eq}$, or equivalently $-T_1 dS_1/dt\ll -T_{\rm int}dS_{\rm
int}/dt$ where $S_{\rm int}$ is the characteristic interior
entropy. The 1.equality between $T_1$ and $\teq$ is
only a very rough estimate guided by present works on atmospheric
models of heavily irradiated planets.}.

Another related problem is the presence of the radiative zone. Again,
the composition is unknown and the opacity data are uncertain in this
relatively high temperature ($T\sim 1500-3000\K$) and high pressure
(up to $\sim 1\,$kbar) regime. Results from our models are based on
opacities from Allard et al. (2001). Other calculations using e.g. the
widely used Alexander \& Ferguson (1994) opacities do yield only a
slightly faster cooling even though the Rosseland opacities are lower
by a factor $\sim 3$ in this regime.

\begin{figure}[htbp]
\hspace*{-1cm}\resizebox{14cm}{!}{\includegraphics[angle=0]{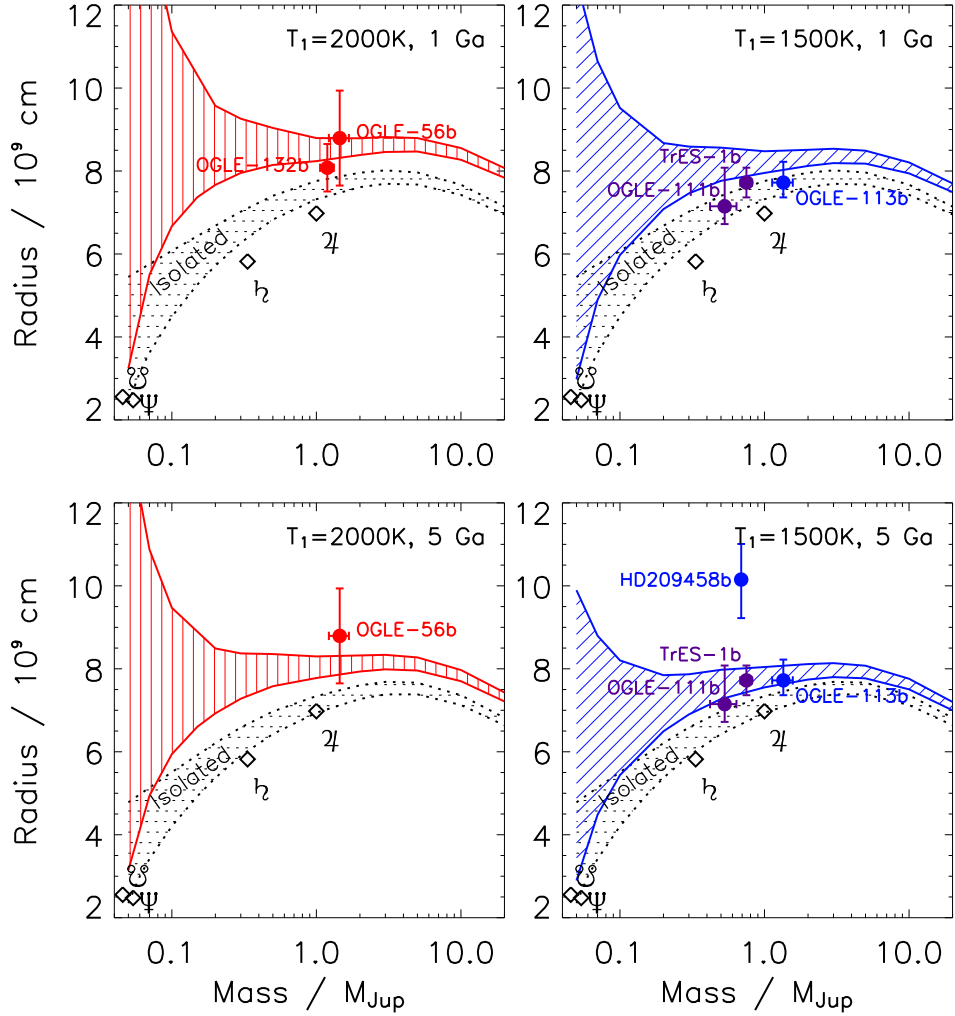}}
%\centerline{\hbox{\psfig{file=poly_index.ps,width=10cm}}}
\caption{Mass-radius relation of strongly irradiated planets with ages
  of 1\,Ga (upper panels) and 5\,Ga (lower panels), and 1-bar temperatures 
  equal to 2000 (left panels) and 1500\,K (right panets),
  respectively. The hashed areas have upper and lower envelopes
  defined by ($Y=0.25$, $M_{\rm core}=0$) and ($Y=0.30$, $M_{\rm
  core}=15\mea$), respectively. Dotted symbols with error bars
  indicate known objects, plotted as a function of their estimated
  1-bar temperatures and ages. Planets whose age is uncertain appear
  in both upper and lower panels. Results for non-irradiated planets
  (dotted lines) are shown for an easier comparison.}
\label{fig:highteq_mr}
\end{figure}

The resulting mass-radius relations are shown in \rfig{highteq_mr} for
$T_1=1500$ and 2000\,K, and compared to the observations for the
planets listed in Table~\ref{tab:transits}. For each
case, an upper limit on the radius is obtained from a pure
hydrogen-helium composition with $Y=0.25$. An ad hoc lower limit comes
from a model with a $15\mea$ central core, and a $Y=0.30$ envelope. In
both case, the opacity table is unchanged.

Figure~\ref{fig:highteq_mr} shows that within uncertainties, the
measurements for 4 planets out of 6 can be explained in the framework
of our simple model. However, two cases stand out: OGLE-TR-132b
appears too small for its age implying that it may contain significant
amounts of heavy elements in a core or in its deep interior.
The case of HD209458b is more problematic: the constraints on
its age, mass, an deep atmospheric temperature that should be $\sim
1500-2000$\,K yield radii that are about 10 to 20\% smaller than measured
(Bodenheimer et al. 2001, 2003; Guillot \& Showman 2002; Baraffe et
al. 2003). The fact that the measured radius corresponds to a
low-pressure ($\sim$mbar) level while the calculated radius
corresponds to a level near 1\,bar is not negligible (Burrows et
al. 2004) but too small to account for the difference. This is
problematic because while it is easy to invoke the presence of a
massive core to explain the small size of a planet, a large size such
as that of HD209458b may require an additional energy source.

Bodenheimer et al. (2001) proposed that this large radius may be
due to a small forced eccentricity ($e\sim 0.03$) of HD209458b, and
subsequent tidal dissipation in the planet interior. In this case,
$\dot\epsilon>0$ in the energy conservation equation (\eref{dldr}).
Because of the relatively limited amount of energy available in the
(non-circular) orbit and the presumably rapid dissipation (due to a
tidal $Q$ that is presumably similar to that of Jupiter, i.e. $Q\sim
10^5-10^6$), this requires the presence of an unseen eccentric
companion. The search for this companion and a possible non-zero
eccentricity of HD209458b is ongoing (Bodenheimer et al. 2003). 

A natural possibility may be the stellar flux itself, since
transporting to deep levels ($\sim 100\,$bars or more) only a small
fraction of order 0.1\% to 1\% of the incoming flux would yield a
radius that is in agreement with the observations. On this basis,
Showman \& Guillot (2002) proposed that kinetic energy generated in
the atmosphere due to the strong asymmetry in stellar insolation may be
transported to deep levels and dissipated there, possibly due to a
small asynchronous rotation and its dissipation by stellar
tides. Another possibility evoked by the authors that km/s atmospheric
winds may maintain the atmosphere into a shear-unstable,
quasi-adiabatic state, which would force temperatures in excess of
3000\,K at levels between 10 and a few tens of bars.

It is puzzling that all other recently announced transiting planets do
not require an additional energy source to explain their size: this is
seen in \rfig{highteq_mr}, which shows that all planets except
HD209458b are consistent with the evolutionary tracks. 

Is there a consistent scenario explaining all the observations? One
possibility is that, as proposed by Bodenheimer et al. (2001),
HD209458b indeed has an eccentric companion. A second possibility is
that their orbital histories have been very different. Finally, the
planets may well have different compositions.

\subsection{\it How do tides and orbital evolution affect the
  contraction and cooling of Pegasi planets?} 

The small orbital eccentricities of Pegasi planets compared to more
distant extrasolar planets tells us that tides raised by the star on
the planet have probably played an important role in circularizing
their orbits, with a timescale estimated at $\sim 1\,$Ga for a planet
at 0.05 AU (Rasio et al. 1996; Marcy et al. 1997). Synchronisation is
expected to occur in only Ma timescales (Guillot et al. 1996), maybe
much less (Lubow et al. 1997). The tides raised by the planet on the
star also tend to spin up the star which leads to a decay of the
planetary orbit. It is interesting to note that, with periods of only
$\sim 1$\,day the three OGLE planets lie close to the orbital
stability threshold (Rasio et al. 1996), or would be predicted to fall
into the star in Ga timescales or less (Witte \& Savonije 2002;
P\"atzold \& Rauer 2002).

The energies available from circularisation and synchronisation 
can be usefully compared to the gravitational energy of the planet
(e.g. Bodenheimer et al. 2001; Showman \& Guillot 2002): 
\begin{eqnarray}
E_{\rm circ}&=&{e^2GM_*M\over a}=
3.6\times 10^{42}\,
\left(e\over 0.1\right)^2\left(M_*\over \msol\right)
\left(M\over\mjup\right)
\left(a\over 10{\rm R_\odot}\right)^{-1}\ \erg, \\
E_{\rm sync} &=& {1\over 2}k^2 MR^2 \Delta\omega^2 =
2.4\times 10^{41}
\left(k^2\over0.25\right)
\left(M\over\mjup\right)
\left(R\over10^{10}{\rm\,cm}\right)^2
\left(\Delta\omega\over 10^{-4}{\rm\,s^{-1}}\right)^2\ \erg, \\
E_{\rm grav}&=& \delta {GM^2\over R} =
2.4\times 10^{42}
\left(\delta\over0.1\right)
\left(M\over\mjup\right)^2
\left(R\over10^{10}{\rm\,cm}\right)^{-1}\ \erg,
\end{eqnarray}
where $e$ is the initial eccentricity, $a$ the planet's orbital distance, $M$
is its mass, $R$ is its radius, $k$ is the dimensionless radius of
gyration, $\Delta\omega$ is the change in the planet's spin before and
after synchronisation, and $\delta$ is approximatively the change in
the planet's radius (neglecting any structural changes in the
calculation of $E_{\rm sync}$ and $E_{\rm grav}$). $E_{\rm grav}$ is
the gravitational energy lost by the planet when its radius decreases by a
factor $\sim\delta$, or alternatively the minimum energy required to
expand its radius by the same factor. 

The fact that the three energy sources are comparable imply that very
early in the evolution, circularisation and synchronisation may have
played a role, perhaps inducing mass loss (Gu et al. 2004). Once
a planet has contracted to a degenerate, low $\theta$ state, the
gravitational energy becomes large, and circularisation and
synchronisation only have a limited role to play. However, two
reservoirs can be invoked: the orbital energy of a massive eccentric
planet that would force a non-zero eccentricity of the inner one
(Bodenheimer et al. 2001) and the absorbed stellar luminosity in its
ability to create kinetic energy in the atmosphere (Showman \& Guillot
2002).

A major uncertainty related to these processes and how they
affect the planetary structure is to know how and where energy is
dissipated. Lubow et al. (1997) proposed that a resonant tidal torque
is exerted at the outer boundary of the inner convection zone, and
that dissipation occurs through the damping of gravity waves
propagating in the outer stable radiative region. Contrary to Jupiter,
this may be an efficient process because Pegasi planets have a
radiative region that extends to great depths. Another possibility is
through the excitation of inertial waves in the convective region, a
process that would occur also in our giant planets (Ogilvie \& Lin
2004). The location of the dissipation is not clear, however. If it
occurs in the atmosphere, the effect of tides on the evolution will be
limited, whereas they will have a maximum impact if they occur deep
into the radiative zone (Guillot \& Showman 2002). 

{\it If\/} dissipation cannot reach into the deep interior, the
planets will not inflate significantly when they migrate to their
present location. This would imply that HD209458b must have migrated
from several AUs to its present location in less than $\sim10\,$Ma
(Burrows et al. 2000b). In this framework, one could invoke a
late migration of the OGLE planets (in particular OGLE-TR-132b) to
explain their relatively small radius compared to HD209458b.

\subsection{\it How does the composition affects the structure and
  evolution?} 

It is generally believed that giant planets of the mass of Jupiter
should have near solar composition and relatively small core
masses. However, it may not be the case: first, Jupiter is in fact
relatively significantly enriched in heavy elements. Second, while
Jupiter is very efficient at ejecting planetesimals from the Solar
System, Pegasi planets are unable to do so because the local orbital
speed $(GM_*/a)^{1/2}\sim 150\rm\,km\,s^{-1}$ is much larger than the
planet's escape velocity $(2GM/R)^{1/2}\sim 50\rm\,km\,s^{-1}$
(Guillot \& Gladman 2000). Furthermore, most planetesimals on low $e$
orbits close to the planet would end up impacting the planet, not the
star (A. Morbidelli, pers. communication 2004). For this reason,
models of in situ formation of Pegasi planets generally yield
large core masses $\sim 40\mea$ (Bodenheimer et al. 2000). 
Pegasi planets should therefore be expected to have very different
compositions and core masses, depending on the properties of the disk
of planetesimals at their formation, the presence of other planets,
and their orbital evolution.

\begin{figure}[htbp]
\hspace*{-1cm}\resizebox{14cm}{!}{\includegraphics[angle=0]{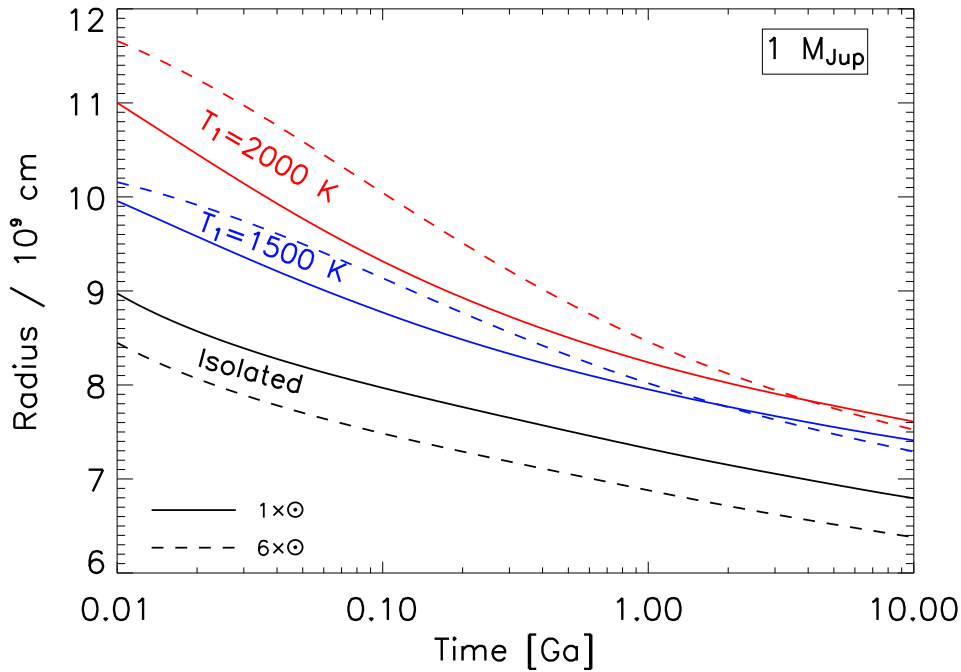}}
%\centerline{\hbox{\psfig{file=poly_index.ps,width=10cm}}}
\caption{Evolution of giant planets in terms of radius vs. time, for
  different irradiation levels, and 2 assumed compositions: solar, and
  6 times solar. (This calculation ignores second order effects as
  modifications of the adiabatic temperature gradient and
  non-linear effects in the opacity calculation, and more importantly
  modifications of atmospheric properties.).}
\label{fig:evols}
\end{figure}

The presence of a core has a relatively straightforward impact on the
evolution of giants planets. As shown in \rfig{highteq_mr}, it leads
to a much faster contraction and a smaller radius at any given age. An
enrichment of the envelope both increases the mean molecular weight
and the opacities, with two opposite effects in terms of the planet's
contraction and cooling. Figure~\ref{fig:evols} shows that for large
irradiations (extended radiative zones), the second effect wins and
leads to a (limited) increase of the planetary radius.  However,
planets with a larger mean molecular weight eventually become smaller.

The difference in inferred radii between HD209458b and other
transiting planets could hence indicate that stellar tides play a role
in slowing or even stalling the contraction of all planets, but that
because of different histories, some planets have a large core mass
but HD209458b has not. In that framework, OGLE-TR-132b would probably
need a core of $\sim 20\,\mea$ or more (or the same amount of
heavy elements in its deep interior) to explain its small radius. The
large [Fe/H] value measured for its parent star
(Table~\ref{tab:transits}) is an indication that the planet may indeed
have grown a large core.

\subsection{{\it What is the role of the atmosphere for the evolution?}}

I have purposedly used a very simple atmospheric model by setting
$T_1\propto\teq=\rm cte$. Of course, this hides many important
complications like opacities, chemistry, gravity dependance, presence
of clouds, atmospheric dynamics, dependance on the incoming stellar
flux...etc. These complications partially explain differences between
several authors (Seager \& Sasselov 1998, 2000; Goukenleuque et
al. 2000; Barman et al. 2001; Sudarsky et al. 2003; Iro et
al. 2004). These works yield characteristic temperatures at the base
of the atmosphere (i.e. where most of the incoming flux has been
absorbed) that range from $\sim 1700$ to $\sim 2300\K$. 

However, the largest differences arise from simple geometrical
reasons: Because these calculations are one-dimensional, some authors
choose to model the atmosphere at the substellar point, some average
the received stellar flux over the day-hemisphere (1/2 less flux), and
others average it over the entire planet (1/4 less flux). This points
to real problems: how does the planet reacts to this extremely
inhomogeneous stellar irradiation, and how do possible inhomogeneities
in the atmosphere affect the planetary evolution?

Without atmospheric dynamics, a synchronous Pegasi planet at $\sim
0.05\,$AU of a G-type star would see its substellar point heated to
$\sim 2500\K$ or more, and its night hemisphere and poles have
temperatures $\sim 100\K$, a clearly unstable situation. Assuming
synchronisation of the convective interior and a radiative atmosphere
obeying the Richardson shear-instability criterion, Showman \& Guillot
(2002) showed that the atmosphere of Pegasi planets are likely to
develop km/s winds, but that spatial photospheric temperature
variations of $\sim 500\K$ are likely. Dynamical models using
shallow-water equations by Cho et al. (2003) also yield latitudinal
temperature variations, but predict a surprising time-dependent
behavior, with a night-side that sometimes becomes hotter than the day
side. A time-dependant approach of radiative transfer, in which the
atmosphere is allowed to react to a varying irradiation, shows that a
km/s rotation indeed yields a $\sim 500\K$ effective temperature
variation. It also shows that the conditions required for the
shallow-water treatment (a relatively long radiative timescale) are
probably not met in Pegasi planets (Iro et al. 2004).

As shown by Guillot \& Showman (2002), to first order (i.e. neglecting
possible non-linear behavior due to e.g. opacity temperature
dependances and/or cloud formation), the cooling with an inhomogeneous
boundary condition is faster than if the same amount of heat has been
homogeneous distributed. This is because heat tends to escape more
rapidly in regions of low atmospheric temperatures. But since the
radiative timescale below optical depth unity is approximatively
$\propto P^2$, levels deeper than a few bars tend to homogeneize
horizontally very efficiently, even with a slow circulation (Iro et
al. 2004). 

Therefore, there is presently no reason to use for evolution models an
atmospheric boundary condition other than that obtained assuming a
stellar flux averaged over the entire planet. Of course, more work is
to be done, as opacity variations, the presence of clouds either on
the day or night side (depending on the kind of circulation),
non-equilibrium chemistry, and possible shear instabilities and
gravity waves damping can all play an important role.

\subsection{{\it Stability and evaporation?}}

Because Pegasi planets are so close to their star, the question of
their survival has been among the first following the discovery of 51
Peg B. Guillot et al. (1996) and Lin et al. (1996) independently
concluded to a relatively fast contraction of the planet and to its
survival based on non-thermal evaporation rates extrapolated from
Jupiter. These evaporation rates $\sim 10^{-16}\msol\,\rm a^{-1}$ turn
out to be extremely close to those inferred from observations of
HD209458b showing the escape of HI (Vidal-Madjar et al. 2003), OI and
CII (Vidal-Madjar et al. 2004). However, the atmospheric escape
problem is more complex than initially envisioned, with
XUV heating, conduction and gravity waves playing important roles
(Lammer et al. 2003; Lecavelier des Etangs et al. 2004).

Generally, a critical question is that of the stability of planets at
close orbital distances in their young ages (Baraffe et al. 2004; Gu
et al. 2004). Figure~\ref{fig:evols} shows that the cooling timescale
is initially relatively long in the case of intense irradiation (see
also fig.~2 of Guillot et al. 1996) and might lead to a significant
mass loss in case of a rapid inward migration because of Roche lobe
overflow (part of the planetary envelope becomes unbound because of
the star's gravitational potential) (Trilling et al. 1998, 2002).
Baraffe et al. (2004) find that another route may be the strong
exospheric evaporation. Below a critical mass, the planet would
inflate before it can become degenerate enough. However, either the
presence of a core and the consequent rapid contraction (see
\rfig{highteq_mr}), or an internal cooling associated to the
decompression upon mass loss may protect the planets from an
exponential evaporation.

\section{Conclusion \& prospects}

We are just beginning to discover the diversity of giant
planets. Already, a variety of problems that are particular to one
planet or a small ensemble of planets have arisen. Given the limited
ensemble of objects that we are given to study and the rapid
evolution of the subject, any attempt to find general rules it fraught
with risk. Some salient conclusions should however resist the trial of
time: 
\begin{itemize}
\item The giant planets of our Solar System all contain a minimum of
  $10\mea$ of heavy elements, and even $\sim 20\mea$ for Saturn and
  probably Jupiter. In Jupiter, most of the heavy elements are mixed
  in the hydrogen-helium envelope. On the contrary, Saturn, Uranus and
  Neptune appear to be significantly differentiated. 
\item The envelopes of Jupiter, Saturn, Uranus and Neptune are
  enriched in heavy elements compared to a solar composition, implying
  that heavy elements were delivered either after the formation
  (requiring large masses in planetesimals because of the low
  accretion probabilities) or when the planets, and in particular
  Jupiter, were not fully formed. In that case, an upward mixing
  (erosion) of these elements with the envelope is required. A third
  possibility is that these elements were captured in an enriched
  nebula.
\item The demixing of helium in metallic hydrogen has probably begun
  in Jupiter, and has been present in Saturn for $2-3$\,Ga. 
\item Like Jupiter and Saturn, the Pegasi planets discovered so far
  are mostly made of hydrogen and helium, but their precise
  composition depends on how tidal effects lead to the dissipation of
  heat in their interior. 
\end {itemize}

Improvements on our knowledge of the giant planets requires a variety
of efforts. Fortunately, nearly all of these are addressed at least
partially by adequate projects in the next few years. The efforts that
are necessary thus include (but are not limited to):
\begin{itemize}
\item Obtain a better EOS of hydrogen, in particular near the
  molecular/metallic transition. This will be addessed by the
  construction of powerful lasers such as the NIF in the US and
  the M\'egaJoule laser in France, and by innovative experiments such
  as shocks on pre-compressed samples. One of the challenges is not only
  obtaining higher pressures, but mostly lower temperatures than
  currently possible with single shocks. The parallel improvement
  of computing facilities should allow more extended numerical
  experiments.
\item Calculate hydrogen-helium and hydrogen-water phase
  diagrams. (Other phase diagrams are desirable too, but of lesser
  immediate importance). This should be possible with new numerical
  experiments. 
\item Have a better yardstick to measure solar and protosolar
  compositions. This may be addressed by the analysis of the Genesis
  mission samples, or may require another future mission. 
\item Improve the values of $J_4$ and $J_6$ for Saturn. This will be
  done as part of the Cassini-Huygens mission. This should lead to
  better constraints, and possibly a determination of whether the
  interior of Saturn rotates as a solid body. 
\item Detect new transiting extrasolar planets, and hopefully some
  that are further from their star. The space missions COROT (2006)
  and {\it Kepler} (2007) should provide the detection and
  characterization of many tens, possibly hundreds of giant planets.
\item Improve the measurement of Jupiter's gravity field, and
  determine the abundance of water in the deep atmosphere. This would
  be possible either from an orbiter, or even with a single fly-by
  (Bolton et al. 2003).
\end{itemize}

Clearly, there is a lot of work on the road, but the prospects for a
much improved knowledge of giant planets and their formation are
bright.

\subsection*{Acknowledgments}

It's a pleasure to thank Didier Saumon, Daniel Gautier, Doug Lin,
Jonathan Fortney, Adam Showman, Bruno B\'ezard, Travis Barman, Brett
Gladman, Chris\-to\-phe Sotin, and Scott Bolton for instructive
discussions, and France Allard, Russ Hemley, Bill Hubbard and
Alessandro Morbidelli for furthermore sharing new or unpublished
results. This work was supported by CNRS and the {\it Programme
National de Plan\'etologie}.

%\end{document}

%----------------------------------------------------------
%\newpage
\section*{References}

%\twocolumn
%\small
%AAAA
\artt{Alexander DR~\& Ferguson JW}{Low-temperature Rosseland opacities}{\apj}{437}{879-91}{1994}
\artt{Allard F, Hauschildt PH, Alexander DR, Tamanai A, \&
Schweitzer A}{The Limiting Effects of Dust in Brown Dwarf Model
  Atmospheres}{\apj}{556}{357-72}{2001}
\artt{Alonso R, Brown TM, Torres G, Latham DW, Sozzetti A, et
  al.}{TrES-1: The Transiting Planet of a Bright K0 V
  Star}{ApJ}{613}{L153}{2004}
\artt{Anders E, Grevesse N}{Abundances of the elements - Meteoritic
  and solar}{Geochem. Cosmo. Acta}{53}{197-214}{1989}
\artt{Anderson JD, Campbell JK, Jacobson RA, Sweetnam DN, Taylor
  AH}{Radio science with Voyager 2 at Uranus - Results on masses and
  densities of the planet and five principal
  satellites}{J. Geophys. Res.}{92}{14877-83}{1987}
\artt{Atreya SK, Mahaffy PR, Niemann HB, Wong MH, Owen TC}{Composition
  and origin of the atmosphere of Jupiter-an update, and implications
  for the extrasolar giant planets}{Plan. Space Sci.}{51}{105-12}{2003}
%BBBB
%\artt{Bahcall JN, Pinsonneault MH}{Solar models with helium and
%heavy elements diffusion.}{Rev. Mod. Phys.}{67}{781}{1995}
\artt{Baraffe I, Chabrier G, Barman TS, Allard F, Hauschildt P}{Evolutionary models for cool brown dwarfs and extrasolar giant planets. The case of HD 209458}{A\&A}{402}{701-12}{2003}
\artt{Baraffe I, Selsis F, Chabrier G, Barman TS, Allard F et
  al.}{The effect of evaporation on the evolution of close-in giant planets}{A\&A}{419}{L13-16}{2004}
\artt{Barman T, Hauschildt PH, Allard F}{Irradiated Planets}{\apj}{556}{885-95}{2001}
\artt{Belov SI, Boriskov GV, Bykov AI, Il'kaev IL, Luk'yanov NB
  et al}{Shock Compression of Solid Deuterium}{Soviet Phys. - JETP
  Lett.}{76}{433-5}{2002} 
\artt{Bodenheimer P, Hubickyj O \& Lissauer JJ}{Models of the in Situ Formation of Detected Extrasolar Giant Planets}{Icarus}{143}{2-14}{2000}
\artt{Bodenheimer P, Lin DNC, Mardling R}{On the Tidal Inflation of Short-Period Extrasolar Planets}{\apj}{548}{466-72}{2001}
\artt{Bodenheimer P, Lin DNC}{Implications of Extrasolar Planets for
  Understanding Planet Formation}{Ann. Rev. Earth
  Plan. Sci.}{30}{113-48}{2002}
\artt{Bodenheimer P, Laughlin G, Lin DNC}{On the Radii of Extrasolar
  Giant Planets}{\apj}{592}{555-63}{2003} 
\artt{Bolton SJ, Allison M, Anderson J, Atreya S, Bagenal F, et al}{A
  Polar Orbiter to Probe Jupiter's Deep Atmosphere, Interior Structure
  and Polar Magnetosphere}{DPS Meeting}{\#35}{\#41.08}{2003}
\artt{Bonev SA, Militzer B, Galli G}{Ab initio simulations of dense
  liquid deuterium: Comparison with gas-gun shock-wave
  experiments}{Phys. Rev. B}{69}{014101}{2004}
\artt{Boss AP}{Possible Rapid Gas Giant Planet Formation in the Solar
  Nebula and Other Protoplanetary Disks}{ApJ}{536}{L101-4}{2000} 
%\artt{Boss AP, Wetherill GW, Haghighipour N}{NOTE: Rapid Formation of
%  Ice Giant Planets}{Icarus}{156}{291-5}{2002}
\artt{Boriskov GV, Bykov AI, Il'kaev IL, Selemir VD, Simakov GV
  et al}{Shock-Wave Compression of Solid Deuterium at a Pressure of
  120 GPa}{Dokl. Phys.}{48}{553-555}{2003}
\artt{Bouchy F, Pont F, Santos NC, Melo C, Mayor M et
  al.}{Two new ``very hot Jupiters'' among the OGLE transiting candidates}{A\&A}{421}{L13-16}{2004}
\artt{Brown TM, Charbonneau D, Gilliland RL, Noyes RW,
Burrows A}{Hubble Space Telescope Time-Series Photometry of the Transiting Planet of HD 209458}{\apj}{552}{699-709}{2001}
\artt{Burrows A, Marley MS \& Sharp, CM}{The Near-Infrared and Optical
  Spectra of Methane Dwarfs and Brown
  Dwarfs}{\apj}{531}{438-446}{2000a} 
\artt{Burrows A, Guillot T, Hubbard WB, Marley MS, Saumon D, et al}{On
  the Radii of Close-in Giant Planets}{\apjl}{534}{L97-100}{2000b}  
\artt{Burrows A, Hubeny I, Hubbard WB, Sudarsky D, Fortney JJ}{Theoretical Radii of Transiting Giant Planets: The Case of OGLE-TR-56b}{\apj}{610}{L53-6}{2004}
%\artt{Busse FH}{Ann. Rev. Fluid. Mech}{10}{435}{1978}
%CCCC
\artt{Campbell JK \& Synnott SP}{Gravity field of the Jovian system from Pioneer and Voyager tracking data}{Astron. J.}{90}{364-72}{1985}
\artt{Campbell JK \& Anderson JD}{Gravity field of the Saturnian system from Pioneer and Voyager tracking data}{Astron. J.}{97}{1485-95}{1989}
\artt{Cavazzoni C, Chiarotti GL, Scandolo S, Tosatti E, Bernasconi
M, \& Parrinello M}{Superionic and Metallic States of Water and Ammonia at Giant Planet Conditions}{Science}{283}{44-46}{1999}
%\artt{Chabrier G., Saumon D., Hubbard W.B, Lunine JI}{The
%molecular-metallic transition of hydrogen and the structure of Jupiter
%and Saturn}{ApJ}{391}{817}{1992}
%\artt{Chabrier G, Barman T, Baraffe I, Allard F, Hauschildt
%  P}{The Evolution of Irradiated Planets: Application to Transits}{\apj}{603}{L53-6}{2004}
\boo{Chandrasekhar S}{Stellar Structure and Evolution}{The University
of Chicago press, Chicago}{1939}  
%\boo{Chandrasekhar S}{Hydrodynamic and hydromagnetic
%stability}{Dover, New York}{1961}
\artt{Charbonneau D, Brown TM, Latham DW, Mayor,
M}{Detection of planetary transits across a Sun-like
star}{\apj}{529}{L45-48}{2000}
\artt{Charbonneau D, Brown TM, Noyes RW, Gilliland RL}{Detection of an
  Extrasolar Planet Atmosphere}{\apj}{568}{377-84}{2002}
\artt{Cho JYK, Menou K, Hansen BMS, Seager S}{The Changing Face of the
  Extrasolar Giant Planet HD 209458b}{ApJ}{587}{L117-20}{2003} 
\artt{Cody AM, Sasselov DD}{HD 209458: Physical Parameters of the
  Parent Star and the Transiting Planet}{ApJ}{569}{451-58}{2002} 
\artt{Cohen ER, Taylor BN}{The 1986 adjustment of the
fundamental physical constants.}{Rev. Mod. Phys.}{59}{1121}{1986}
\artt{Collins GW, Da Silva LB, Celliers P, et al}{Measurements of the equation of state of deuterium at the fluid insulator-metal transition}{Science}{281}{1178-81}{1998}
\artt{Conrath BJ~\& Gautier D}{Saturn Helium Abundance: A Reanalysis of Voyager Measurements}{Icarus}{144}{124-34}{2000}
%DDDD
\artt{da Silva LB, Celliers P, Collins GW, Budil KS, Holmes NC
  et al}{Absolute Equation of State Measurements on Shocked Liquid
  Deuterium up to 200 GPa (2 Mbar)}{Phys. Rev. Lett.}{78}{483-6}{1997}
\artt{Datchi F, Loubeyre P, Letoullec R}{Extended and accurate
  determination of the melting curves of argon, helium, ice (H2O), and
  hydrogen (H2)}{Phys. Rev. B}{61}{6535-46}{2000} 
\artt{Davies ME, Abalakin VK, Bursa M, Lederle T, Lieske JH et
  al}{Report of the IAUIAG COSPAR working group on cartographic
  coordinates and rotational elements of the planets and satellites:
  1985}{Celestial Mech.}{39}{102-13}{1986} 
\artt{Desjarlais MP}{Density-functional calculations of the liquid
  deuterium Hugoniot, reshock, and reverberation
  timing}{Phys. Rev. B}{68}{064204}{2003}
%EEEE
%FFFF
%\art{Fegley BJ~\& Lodders K}{Icarus}{110}{117}{1994}
%\art{Folkner WM, Woo R, Nandi S}{\JGR}{103}{22831}{1998}
\artt{Fortney J, Hubbard WB}{Phase separation in giant planets: inhomogeneous evolution of Saturn}{Icarus}{164}{228-43}{2003}
%\artt{Fortney J, Hubbard WB}{Effects of Helium Phase Separation on the Evolution of Extrasolar Giant Planets}{ApJ}{608}{1039-49}{2004}
%GGGG
%\artt{Galli G. et al}{Phys. Rev. B}{61}{909}{2000}
\bib{Gautier D, Owen, T}. 1989. The composition of outer planet
atmospheres. In {\it Origin and Evolution of Planetary and Satellite
  Atmospheres}, ed. SK Atreya, JB Pollack, MS Matthews, 487--512. University
of Arizona Press: Tucson
\artt{Gautier D, Hersant F, Mousis O \& Lunine JI}{Enrichments in
  Volatiles in Jupiter: A New Interpretation of the Galileo
  Measurements}{\apjl}{550}{L227-230}{2001} 
%\artt{Goldreich P \& Soter S}{Icarus}{5}{375}{1966}
\artt{Goukenleuque C, B\'ezard B, Joguet B, Lellouch E, Freedman
R}{A radiative equilibrium model of 51 Peg b}{\icarus}{143}{308}{2000}
%\artt{Graboske HC, Pollack JB, Grossman AS \& Olness
%RJ}{\apj}{199}{265}{1975}
\artt{Gregoryanz E, Goncharov AF, Matsuishi K, Mao H, Hemley
  RJ}{Raman Spectroscopy of Hot Dense Hydrogen}{Phys Rev. Lett.}{90}{175701-1-4}{2003}
%\artt{Gudkova TV~\& Zharkov VN}{\planss}{47}{1201}{1999}
\artt{Gu P-G, Bodenheimer PH, Lin DNC}{The Internal Structural
  Adjustment Due to Tidal Heating of Short-Period Inflated Giant
  Planets}{\apj}{608}{1076-94}{2004}
\artt{Guillot T, Gautier D, Chabrier G, Mosser
B}{Are the giant planets fully convective?}{Icarus}{112}{337-53}{1994}
\artt{Guillot T, Chabrier G, Gautier D \& Morel P}{Effect of radiative
  transport on the evolution of Jupiter and
  Saturn}{ApJ}{450}{463-72}{1995}
%\artt{Guillot T \& Morel P}{A\&AS}{109}{109}{1995}
\artt{Guillot T}{Condensation of Methane Ammonia and Water in the Inhibition of Convection in Giant Planets}{Science}{269}{1697-99}{1995}
\artt{Guillot T, Burrows A, Hubbard WB, Lunine JI \& Saumon
D}{Giant planets at small orbital distances}{\apj}{459}{L35-38}{1996}
%\artt{Guillot T, Gautier D \& Hubbard WB}{Icarus}{130}{534}{1997}
\artt{Guillot T}{A comparison of the interiors of Jupiter and
  Saturn}{\planss}{47}{1183-200}{1999a} 
\artt{Guillot T}{Interior of Giant Planets Inside and Outside the
  Solar System}{Science}{286}{72-77}{1999b} 
\bib Guillot T Gladman B. 2000. Late Planetesimal Delivery and the
  Composition of Giant Planets. In {\em Proceedings of the Disks,  
  Planetesimals and Planets Conference, ASP Conference Series},
  eds F Garzon et al., 219:475-485.
\artt{Guillot T \& Showman A}{Evolution of ``51 Pegasus b-like''
  planets}{A\&A}{385}{156-65}{2002} 
\boa{Guillot T, Stevenson DJ, Hubbard WB, Saumon D}{The interior of
  Jupiter}{Jupiter: The Planet, Satellites, and Magnetosphere}{ed.  F
  Bagenal, W McKinnon, T Dowling, in press}{2004}
%HHHH
\artt{Henry GW, Marcy GW, Butler RP, Vogt,
SS}{A transiting ``51 Peg-like'' planet}{\apj}{529}{L41-44}{2000}
\artt{Hersant F, Gautier D, Lunine JI}{Enrichment in volatiles in the
  giant planets of the Solar System}{Plan. Space
  Sci.}{52}{623-41}{2004} 
\artt{Holmes NC, Ross M, Nellis WJ}{Temperature measurements and
  dissociation of shock-compressed liquid deuterium and
  hydrogen}{Phys. Rev. B}{52}{15835-45}{1995}
\artt{Hubbard WB}{Thermal structure of Jupiter.}{\ApJ}{152}{745-54}{1968} 
%\artt{Hubbard WB~\& Lampe M}{\apjs}{18}{297}{1969}
\artt{Hubbard WB}{The Jovian surface condition and cooling
  rate}{Icarus}{30}{305-10}{1977}
\artt{Hubbard WB}{Effects of differential rotation on the
gravitational figures of Jupiter and Saturn.}{Icarus}{52}{509-15}{1982}
%\boo{Hubbard WB}{Planetary Interiors}{Van Nostrand
%Reinhold Co., Inc, New York}{1984}
%\bib{Hubbard, WB} In {\it Origin and Evolution of Planetary and Satellite 
%Atmospheres}, S K Atreya, J B Pollack, and M S Matthews, eds.,
%University of Arizona Press, Tucson, pp. 539--563 (1989)
%\artt{Hubbard WB \& Marley MS}{Icarus}{78}{102}{1989}
\bib Hubbard WB, Pearl JC, Podolak M, Stevenson DJ. 1995. The Interior
of Neptune. In {\it Neptune and Triton}, ed. DP
Cruikshank. 109--138. Univ. of Arizona Press: Tucson
\artt{Hubbard WB, Guillot T, Marley MS, Burrows A, Lunine JI \& Saumon
DS}{Comparative evolution of Jupiter and Saturn}{\planss}{47}{1175-82}{1999}
\artt{Hubbard WB}{Gravitational signature of Jupiter's deep
zonal flows.}{Icarus}{137}{196-99}{1999}
\artt{Hubbard WB, Burrows A, Lunine JI}{Theory of Giant
  Planets}{Ann. Rev. Astron. Astrophys.}{40}{103-36}{2002}
%\artt{Hubbard WB, Fortney JJ, Lunine JI, Burrows A, Sudarsky
%D \& Pinto P}{\apj}{560}{413}{2001}
%IIII
%\artt{Ingersoll AP, Kanamori H \& Dowling
%TE}{Waves from the Collisions of Comet Shoemaker-Levy-9 with Jupiter}{Geophys. Res. Lett.}{21}{1083}{1994}
%\bib {Ingersoll AP, Barnet CD, Beebe RF et al}, in {\it Neptune
%and Triton}, ed D.P Cruikshank, University of Arizona Press, Tucson,
%613 (1995)
\bib{Iro N, B\'ezard B, Guillot T}. 2004. A Time-dependent radiative
model of HD209458b. {submitted to Icarus}
%JJJJ
%\artt{Jeffreys H.}{M.N.R.A.S.}{83}{350}{1923}
%KKKK
%\boo{Kippenhahn R \& Weigert A}{Stellar structure and
%evolution}{Springer-Verlag, Berlin}{1991}
\artt{Konacki M, Torres G, Jha S, Sasselov
  DD}{An extrasolar planet that transits the disk of its parent
  star}{Nature}{421}{507-9}{2003}
\artt{Konacki M, Torres G, Sasselov DD, Pietrzynski G, Udalski A et
  al.}{The Transiting Extrasolar Giant Planet around the Star
  OGLE-TR-113}{ApJ}{609}{L37-40}{2004}
\artt{Knudson MD, Hanson DL, Bailey JE, Hall CA, Asay JR, Deeney
  C}{Principal Hugoniot, reverberating wave, and mechanical reshock
  measurements of liquid deuterium to 400 GPa using plate impact
  techniques}{Phys. Rev. B}{69}{144209}{2004}
\artt{Knudson MD, Hanson DL, Bailey JE, Hall CA, Asay JR, Anderson
  WW}{Equation of State Measurements in Liquid Deuterium to 70
  GPa}{Phys. Rev. Lett.}{87}{225501}{2002}
%LLLL
\artt{Lammer H, Selsis F, Ribas I, Guinan EF, Bauer SJ, Weiss
  WW}{Atmospheric Loss of Exoplanets Resulting from Stellar X-Ray and
  Extreme-Ultraviolet Heating}{\apj}{598}{L121-4}{2003}
\artt{Laughlin G, Wolf A, Vanmunster T, Bodenheimer P, Fischer D et
  al.}{A Comparison of Observationally Determined Radii with
  Theoretical Radius Predictions for Short-Period Transiting
  Extrasolar Planets}{\apj}{}{submitted}{2004}
\artt{Lecavelier des Etangs A, Vidal-Madjar A, McConnell JC, H\'ebrard
  G}{Atmospheric escape from hot Jupiters}{A\&A}{418}{L1-4}{2004}
\artt{Lenzuni P, Chernoff DF, Salpeter EE}{Rosseland and Planck mean
  opacities of a zero-metallicity gas}{\apjs}{76}{759-801}{1991} 
\artt{Levison HF, Morbidelli A}{The formation of the Kuiper belt by
  the outward transport of bodies during Neptune's
  migration}{Nature}{426}{419-21}{2003} 
\artt{Lin DNC, Bodenheimer P, \& Richardson DC}{Orbital migration of the planetary companion of 51 Pegasi to its present location}{\nat}{380}{606-7}{1996} 
\artt{Lindal GF, Wood GE, Levy GS, Anderson JD, Sweetnam DN, et al}{The atmosphere of Jupiter - an analysis of the Voyager radio occultation measurements}{J. Geophys. Res.}{86}{8721-7}{1981}
\artt{Lindal GF, Sweetnam DN, Eshleman VR}{The atmosphere of
Saturn -- an analysis of the Voyager radio occultation
  measurements}{Astron. J.}{90}{1136-46}{1985} 
\artt{Lindal GF}{The atmosphere of Neptune -- an analysis of radio occultation data acquired with Voyager 2}{Astron. J.}{103}{967-82}{1992}
\artt{Lodders K}{Solar System Abundances and Condensation Temperatures of the Elements}{ApJ}{591}{1220-47}{2003}
\artt{Lubow SH, Tout CA \& Livio M}{Resonant Tides in Close Orbiting Planets}{\apj}{484}{866-70}{1997}
%MMMM
\artt{Magalh\~aes JA, Seiff A, Young RE}{The Stratification of
  Jupiter's Troposphere at the Galileo Probe Entry
  Site}{Icarus}{158}{410-33}{2002}
\artt{Mahaffy PR, Niemann HB, Alpert A, Atreya SK, Demick J et
  al.}{Noble gas abundance and isotope ratios in the atmosphere of Jupiter from the Galileo Probe Mass Spectrometer}{JGR}{105}{15061-72}{2000}
\artt{Mao H, Hemley RJ}{Ultrahigh-pressure transitions in solid
  hydrogen}{Rev. Mod. Phys.}{66}{671-92}{1994} 
\artt{Marcy GW, Butler RP, Williams E, Bildsten L, Graham JR,
Ghez AM, Jernigan JG}{The Planet around 51 Pegasi}{ApJ}{481}{926-35}{1997}
%\bib Marcy GW, Cochran WD \& Mayor M In {\it Protostars and
%Planets IV} (Tucson: University of Arizona Press; eds Mannings V.,
%Boss A.P, Russell SS), p. 1285 (2000)
\artt{Marley MS, Gomez P \& Podolak P}{Monte Carlo interior models for
  Uranus and Neptune}{J. Geophys. Res.}{100}{23349-54}{1995}
\artt{Marley MS~\& McKay CP}{Thermal Structure of Uranus' Atmosphere}{Icarus}{138}{268-86}{1999}
%\artt{Mayor M \& Queloz D}{A Jupiter-mass companion to a
%solar-type star}{\nature}{378}{355}{1995}
\artt{Militzer B \& Ceperley DM}{Path Integral Monte Carlo
Simulation of the Low-Density Hydrogen
Plasma}{Phys. Rev. E}{63}{6404}{2001}
\artt{Moutou C, Pont F, Bouchy F \& Mayor M}{Accurate radius and mass
  of the transiting exoplanet OGLE-TR-132b}{A\&A}{424}{L31}{2004}
%NNNN
%\artt{Nellis WJ, Weir ST, Mitchell AC}{Phys. Rev. B}{59}{3434}{1999}
\artt{Niemann HB, Atreya SK, Carignan GR, Donahue TM,
Haberman JA, et al}{The composition of
the jovian atmosphere as determined by the Galileo probe mass
spectrometer.}{\JGR}{103}{22831-8}{1998}
%OOOO
\artt{Ogilvie GI, Lin DNC}{Tidal Dissipation in Rotating Giant
  Planets}{ApJ}{610}{477-509}{2004}
\artt{Owen T, Mahaffy P, Niemann HB, Atreya S, Donahue T, Bar-Nun A \& de Pater I}{A low-temperature origin for the planetesimals that formed Jupiter}{\nat}{402}{269-70}{1999}
%PPPP
%\artt{de\,Pater I, Massie, ST}{Icarus}{62}{143}{1985}
\artt{P\"atzold M, Rauer H}{Where Are the Massive Close-in Extrasolar
  Planets?}{\apj}{568}{L117-20}{2002}
\artt{Pearl JC \& Conrath BJ}{The albedo, effective temperature,
and energy balance of Neptune, as determined from Voyager
data.}{J. Geophys. Res. Suppl.}{96}{18921-9}{1991}
\artt{Pfaffenzeller O, Hohl D \& Ballone P}{Miscibility of hydrogen
and helium under astrophysical conditions}{Phys Rev. Lett.}{74}{2599-602}{1995}
\bib Podolak M, Hubbard WB, Stevenson DJ. 1991. Model of Uranus' interior
  and magnetic field. In {\it Uranus}, ed. JT Bergstralh, ED Miner,
  MS Matthews. 29--61. Univ. of Arizona Press: Tucson
\artt{Podolak M, Weizman A \& Marley MS}{Comparative models of Uranus
  and Neptune}{\pss}{43}{1517-22}{1995}
\artt{Podolak M, Podolak JI \& Marley MS}{Further investigations of
  random models of Uranus and Neptune}{\planss}{48}{143-51}{2000} 
\artt{Pollack JB, Hubickyj O, Bodenheimer P, Lissauer JJ, Podolak
M \& Greenzweig Y}{Formation of the Giant Planets by Concurrent
  Accretion of Solids and Gas}{Icarus}{124}{62}{1996} 
\artt{Pont F, Bouchy F, Queloz D, Santos NC, Melo C, Mayor M, Udry
  S}{The ``missing link'': A 4-day period transiting exoplanet around
  OGLE-TR-111}{A\&A}{426}{L15}{2004}
%QQQQ

%RRRR
%\artt{Rogers FJ}{Physics of Plasmas}{7}{51}{2000}
\artt{Rasio F, Tout CA, Lubow SH, Livio M}{Tidal Decay of Close
  Planetary Orbits}{\apj}{470}{1187-91}{1996}
\artt{Ross M}{Linear-mixing model for shock-compressed liquid
  deuterium}{Phys Rev. B.}{58}{669}{1998} 
\artt{Ross M, Yang LH}{Effect of chainlike structures on
  shock-compressed liquid deuterium}{Phys. Rev. B}{64}{134210}{2001}
\artt{Roos-Serote M, Atreya SK, Wong MK, Drossart P}{On the water
  abundance in the atmosphere of Jupiter}{Plan. Space
  Sci.}{52}{397-414}{2004} 
\bib{Roulston, MS, Stevenson, DJ}. 1995. Prediction of neon depletion in
Jupiter's atmosphere. {\it EOS} {\bf 76}: 343 [abstract]
%SSSS
%\artt{Salpeter EE}{ApJ}{181}{L83}{1973}
\artt{Sasselov DD}{The new transiting planet OGLE-TR-56b: Orbit and
  atmosphere}{\apj}{596}{1327-31}{2003}
\artt{Saumon D, Hubbard WB, Chabrier G, Van Horn HM.}{The role of the
  molecular-metallic transition of hydrogen in the evolution of Jupiter,
  Saturn and brown dwards.}{ApJ}{391}{827-31}{1992} 
\artt{Saumon D, Chabrier G \& Van Horn HM}{An equation
of state for low-mass stars and giant planets.}{\apjs}{99}{713-41}{1995}
\artt{Saumon D, Hubbard WB, Burrows A, Guillot T, Lunine JI \&
Chabrier G}{A Theory of Extrasolar Giant Planets}{\apj}{460}{993-1018}{1996}
\artt{Saumon D, Chabrier G, Wagner DJ, \& Xie X}{Modeling
  Pressure-Ionization of Hydrogen in the Context of Astrophysics}{High
  Pressure Research}{16}{331-43}{2000} 
\artt{Saumon D, Guillot T}{Shock Compression of Deuterium and the
  Interiors of Jupiter and Saturn}{ApJ}{609}{1170-80}{2004}
\artt{Seager S \& Sasselov DD}{Extrasolar giant planets
  under strong stellar irradiation}{\apj}{502}{L157-60}{1998}
\artt{Seager S \& Sasselov DD}{Theoretical Transmission Spectra during
  Extrasolar Giant Planet Transits}{\apj}{537}{916-21}{2000} 
\artt{Seager S \& Hui L}{Constraining the Rotation Rate of Transiting
  Extrasolar Planets by Oblateness Measurements}{\apj}{574}{1004-10}{2002}
\artt{Seiff A, Kirk DB, Knight TCD, Young RE, Mihalov JD et al}
     {Thermal structure of Jupiter's atmosphere near the
       edge of a 5-$\mu$m hot spot in the north equatorial
       belt.}{\JGR}{103}{22857-90}{1998}
\artt{Showman AP \& Guillot T}{Atmospheric circulation and tides of
  ``51 Pegasus b-like'' planets}{A\&A}{385}{166-80}{2002} 
\bib{Sozzetti A, Young D, Torres G, Charbonneau D, Latham DW, et
  al. 2004, High-Resolution Spectroscopy of the Transiting Planet Host
  Star TrES-1, ApJL, in press (astro-ph/0410483)}
\artt{Stevenson DJ \& Salpeter EE}{The dynamics and helium
  distribution in hydrogen-helium fluid planets}{ApJ Suppl}{35}{239-61}{1977}
%\artt{Stevenson DJ \& Salpeter EE}{ApJ Suppl.}{35}{239}{1977b}
\artt{Stevenson DJ}{Interiors of the giant planets}{Ann. Rev. Earth Planet. Sci.}{10}{257-95}{1982}
%\artt{Stevenson DJ}{Planetary Magnetic
%fields}{Rep Prog. Phys.}{46}{555}{1983}
%\artt{Stevenson D.J.}{\araa}{29}{163}{1991}
\artt{Stevenson DJ}{Cosmochemistry and structure of the giant planets
  and their satellites}{Icarus}{62}{4-15}{1985} 
\artt{Sudarsky D, Burrows A, Hubeny I}{Theoretical Spectra and
  Atmospheres of Extrasolar Giant Planets}{\apj}{588}{1121-48}{2003} 
%TTTT
\artt{Torres G, Konacki M, Sasselov DD, Jha S}{New Data and Improved
  Parameters for the Extrasolar Transiting Planet
  OGLE-TR-56b}{ApJ}{609}{1071--5}{2004} 
\artt{Trilling DE, Benz W, Guillot T, Lunine JI, Hubbard WB, \&
  Burrows A}{Orbital Evolution and Migration of Giant Planets:
  Modeling Extrasolar Planets}{\apj}{500}{428-39}{1998}
\artt{Trilling DE, Lunine JI, Benz W}{Orbital migration and the
  frequency of giant planet formation}{A\&A}{394}{241-51}{2002}
\artt{Tyler GL, Sweetnam DN, Anderson JD,  Borutzki SE, Campbell JK,
  et al}{Voyager radio science observations of Neptune and
  Triton}{Science}{246}{1466-73}{1989} 
%UUUU
%VVVV
\artt{Vidal-Madjar A, D\'esert J-M, Lecavelier des Etangs A, H\'ebrard
  G, Ballester GE et al}{Detection of Oxygen and Carbon in the
  Hydrodynamically Escaping Atmosphere of the Extrasolar Planet HD
  209458b}{\apj}{604}{L69-72}{2004} 
\artt{Vidal-Madjar A, Lecavelier des Etangs A, D\'esert J-M, Ballester
  GE, Ferlet R, et al}{An extended upper atmosphere around the
  extrasolar planet HD209458b}{Nature}{422}{143-6}{2003} 

%WWWW
\artt{Warwick JW, Evans DR, Roming JH, Sawyer CB, Desch MD,
  et al}{Voyager 2 radio observations of
  Uranus}{Science}{233}{102-6}{1986} 
\artt{Warwick JW, Evans DR, Peltzer GR, Peltzer RG, Roming JH,
  et al}{Voyager planetary radio astronomy at
  Neptune}{Science}{246}{1498-501}{1989} 
\artt{Weir ST, Mitchell AC, Nellis WJ}{Metallization of Fluid
  Molecular Hydrogen at 140 GPa (1.4
  Mbar)}{Phys. Rev. Lett.}{76}{1860-3}{1996}
\artt{Witte MG, Savonije GJ}{Orbital evolution by dynamical tides in
  solar type stars. Application to binary stars and planetary
  orbits}{A\&A}{386}{222-36}{2002}
%XXXX
% YYYY
%ZZZZ
\artt{von Zahn U, Hunten DM, Lehmacher G}{Helium in
Jupiter's atmosphere: results from the Galileo probe helium
interferometer experiment}{\JGR}{103}{22815-30}{1998}
\artt{Zharkov VN \& Trubitsyn VP}{Determination of the equation of
  state of the molecular envelopes of Jupiter and Saturn from their
  gravitational moments}{Icarus}{21}{152-6}{1974}
\boo{Zharkov VN \& Trubitsyn VP}{Physics of Planetary
Interiors}{Ed. WB Hubbard, Pachart:Tucson}{1978}
%\boo{Zharkov V.N.}{Interior structure of the Earth and
%planets}{Harwood academic pubishers}{1986}
%\boa{Zharkov V.N. et Gudkova T.V.}{High pressure research:
%application to Earth and planetary sciences}{(eds. Y. Syono et M.H.
%Manghnani), TERRAPUB, Tokyo}{1992}

%\end{thebibliography}

\end{document}